\documentclass[prc,aps,amsmath,amssymb,superscriptaddress,twocolumn,showpacs,floatfix,a4paper]{revtex4-2}

\usepackage{graphicx,colordvi}
\usepackage{dcolumn}
\usepackage{bm}
\usepackage{threeparttable}
\usepackage{xspace}
\usepackage{gensymb}
\usepackage{cases}
\usepackage{textcomp}
\usepackage{tabularx,booktabs}
\usepackage{epstopdf}
\usepackage{physics} 
\usepackage{mathtools, amssymb, amsthm, amsmath}

\usepackage{appendix}
\usepackage{csquotes}

\usepackage{xcolor}
\usepackage{tcolorbox}

\usepackage[dvipsnames]{xcolor}
\usepackage{bigints}
\usepackage{scalerel}

\begin{document}

\title{Microscopic description of the fission process including intrinsic excitations \\ Part III: $^{240}$Pu fission dynamics along 1D asymmetric paths within the Schr\"odinger Collective Intrinsic Model}

\author{P. Carpentier}
\affiliation{%
 CEA, DAM, DIF, F-91297 Arpajon cedex, France
}%
\affiliation{%
  Universit\'e Paris-Saclay, CEA, LMCE, 91680, Bruyères-le-Châtel, France
}%

\author{N. Pillet}%
\affiliation{%
 CEA, DAM, DIF, F-91297 Arpajon cedex, France
}%
\affiliation{%
  Universit\'e Paris-Saclay, CEA, LMCE, 91680, Bruyères-le-Châtel, France
}%

\author{R. Bernard}%
\affiliation{%
CEA, DES, IRESNE, DER, SPRC, LEPh, 13115 Saint-Paul-lès-Durance, France
}%

\author{L.M. Robledo}%
\affiliation{%
Center for Computational Simulation, Universidad Polit\'ecnica de 
Madrid, Campus Montegancedo, 28660 Boadilla del Monte, Madrid, Spain
}%
\affiliation{Departamento  de F\'{\i}sica Te\'orica and CIAFF, 
Universidad Aut\'onoma de Madrid, 28049-Madrid, Spain}%

\author{D. Lacroix}%
\affiliation{%
Université Paris-Saclay, CNRS/IN2P3, IJCLab, Orsay, 91405, France
}%

\author{N. Dubray}%
\affiliation{%
 CEA, DAM, DIF, F-91297 Arpajon cedex, France
}%
\affiliation{%
  Universit\'e Paris-Saclay, CEA, LMCE, 91680, Bruyères-le-Châtel, France
}%

\author{D. Regnier}%
\affiliation{%
 CEA, DAM, DIF, F-91297 Arpajon cedex, France
}%
\affiliation{%
  Universit\'e Paris-Saclay, CEA, LMCE, 91680, Bruyères-le-Châtel, France
}%

\author{W. Younes}%
\affiliation{%
Nuclear Science Division, Lawrence Berkeley National Laboratory, Berkeley, California 94720, USA
}%

\date{\today}

\begin{abstract}
This last article of the trilogy \cite{trilogy1,trilogy2,trilogy3} focuses on the dynamical equation of the Schr\"odinger Collective-Intrinsic Model (SCIM). 
First, we motivate and discuss the need to regularize the adiabatic and excited dynamical ingredients entering the collective-intrinsic Hamiltonian, namely the collective potential, the collective inertia tensor, and the collective dissipative tensor. 
In particular, we introduce a Savitzky-Golay low-pass filter to remove numerical fluctuations incompatible with the second-order truncation in the Symmetric Ordered Product of Operators used to derive the SCIM equations. 
The diagonal and off-diagonal properties of the three dynamical ingredients are then analyzed along the asymmetric fission path in $^{240}$Pu. 
This study highlights the dominant role of neutron and proton excitation channels, especially in the second well and scission regions, whereas proton-neutron couplings remain essentially negligible. Furthermore, in the adiabatic limit of the SCIM, we perform a comparison with the Gaussian Overlap approximation for both zero point energies and masses, which reveals very close predictions.
Second, we discuss the construction of the initial wave packet and the numerical resolution of the collective-intrinsic Schr\"odinger equation. 
Using a continuity equation, we derive the probability fluxes associated with the different components of the wave function, which provide direct access to the contribution of the different excitations to the final observables for the fission problem. 
The excited states are found to account for more than 80\% of the total flux at scission. 
Finally, we evaluate, within the SCIM framework, the neutron and proton fragment distributions as well as the energy balance, including the total kinetic and excitation energies. 
The obtained results are found to be consistent with available experimental data and demonstrate the importance of explicitly including intrinsic excitations in the description of fission dynamics.
\end{abstract}

\maketitle

\section{Introduction}

In the first two articles of this trilogy, we introduced three new protocols, the \enquote{Link}, \enquote{Drop}, and \enquote{Continuous Deflation} procedures, to construct potential energy surfaces (PES) with continuity and regularity properties compatible with the Schrödinger Collective Intrinsic Model (SCIM). The SCIM extends the Time-Dependent Generator Coordinate Method (TDGCM) by incorporating selected classes of intrinsic excitations through a second-order expansion based on Symmetric Ordered Products of Operators (SOPO) \cite{rsbook}. This formalism transforms the original non-local problem into a local Schrödinger-like equation governed by a collective Hamiltonian, while absorbing the non-locality into generalized collective quantities.

The SCIM collective-intrinsic Hamiltonian can be written as
\begin{eqnarray}\label{hscim}
\mathcal{H}_{SCIM}(\bar q) = V_{{SCIM}}(\bar q) + [D_{{SCIM}}(\bar q)\frac{\partial}{\partial q}]^{(1)} \nonumber \\ +
[B_{{SCIM}}(\bar q)\frac{\partial}{\partial q}]^{(2)}, 
\end{eqnarray}
where $V_{SCIM}(\bar q)$, $D_{SCIM}(\bar q)$, and $B_{SCIM}(\bar q)$ denote the collective potential, dissipative tensor, and inertia tensor, respectively. These quantities are evaluated at the center of mass $\bar q=(q+q')/2$ of the collective coordinates. Their explicit expressions are given in Appendix C of the first article of the trilogy \cite{trilogy1}. Throughout this work, they are computed using the Gogny D1S interaction; additional details can be found in Ref.~\cite{TPaul}.

The primary objective of this third article is to perform the dynamical propagation of a collective wave packet including both the adiabatic $\tilde{\mathcal{P}}_{20}$ path and the Continuous Deflation excited paths, coupled through the SCIM Hamiltonian.
Section \ref{dyna} presents the adiabatic collective ingredients required for the dynamical calculations. We first discuss the Savitzky--Golay regularization used to ensure the smooth second-order variations required by the SCIM formalism. We then establish a formal correspondence between the SCIM and the Gaussian Overlap Approximation (GOA), with particular emphasis on the comparison of zero-point energies and collective inertias. The resulting SCIM collective potential, inertia tensor, and dissipative tensor are subsequently analyzed.
Section \ref{dynaing} extends this analysis to the excited quantities by examining the diagonal and off-diagonal collective couplings generated by the SCIM Hamiltonian. Finally, Section \ref{propagat} presents the propagation of the collective wave packet. We describe the extraction of particle fluxes, which provides access to fragment properties at scission in the presence of intrinsic excitations. Particular attention is devoted to particle-number distributions and to the corresponding energy balance. Conclusions and perspectives are given in Section \ref{conclu}.

\section{Adiabatic collective dynamical ingredients}\label{dyna}

In this section, we investigate the collective dynamical ingredients of the SCIM in the adiabatic limit, where intrinsic excitations are neglected.
We first discuss the extraction of the collective potential and inertia tensor, introducing the well-established Savitzky-Golay filter to regularize high-frequency fluctuations. The purpose of this smoothing procedure is to stabilize the evaluation of the SCIM Hamiltonian while preserving its physically relevant low-frequency content.
We then establish a formal correspondence between the SCIM and the Gaussian Overlap Approximation (GOA). After briefly recalling the main features of the GOA, we use it throughout this section as a reference framework to assess the validity of the SCIM in the adiabatic limit.
Finally, we compare the collective ingredients obtained within the SCIM with those predicted by the exact GOA and by the conventional GOA+Cranking and GOA+ATDHFB local approximations. Particular emphasis is placed on the role of non-local effects in the description of collective dynamics.

\subsection{Savitzky-Golay regularization of SCIM dynamical ingredients}\label{savgol}

The extraction of the SCIM dynamical ingredients relies on the inversion of the norm kernel $\bar{\mathcal N}$ entering the generalized Hill-Wheeler equation (see Eq. (14) of the first article of this trilogy \cite{trilogy1}). The operator $\bar{\mathcal N}$ takes the form of a differential operator containing both derivative and non-derivative terms built from the moments of the overlap kernel. Its inversion is therefore non-trivial and requires a dedicated iterative factorization procedure, described in Appendix B of Ref. \cite{trilogy1}. This procedure combines successive factorizations of the zeroth-order overlap moment $\mathcal N^{(0)}$ with a truncated expansion of derivative operators up to second order. For the iterative inversion to converge, higher-order derivatives of the overlap moments must rapidly become negligible. This requirement translates into a sufficient degree of regularity of the overlap moments along the collective path. 

The inversion yields the inverse square-root operator $\bar{\mathcal N}^{-1/2}$ entering the definition of the SCIM collective Hamiltonian:
\begin{equation}
\mathcal H_{{SCIM}}
=
\bar{\mathcal N}^{-1/2}
\bar{\mathcal H}
\bar{\mathcal N}^{-1/2}
\end{equation}
whose explicit evaluation requires repeated SOPO products truncated at second order, as detailed in Appendix C of the first article of the trilogy \cite{trilogy1}. The validity of this second-order truncation likewise relies on sufficient regularity of both the overlap and Hamiltonian moments. Otherwise, the truncated expansion fails to provide a reliable approximation, leading to nonphysical collective tensors. The SCIM Hamiltonian takes the canonical form in terms of the collective coordinate $c_\#$:
\begin{align}
\mathcal{H}_{{SCIM}}(c_\#)
=
V_{{SCIM}}(c_\#) + \left[
D_{{SCIM}}(c_\#)
\frac{\partial}{\partial c_\#}
\right]^{(1)}
\nonumber \\  \; \; \;+
\left[
B_{{SCIM}}(c_\#)
\frac{\partial}{\partial c_\#}
\right]^{(2)}.
\end{align}

In the adiabatic limit, the SCIM Hamiltonian reduces to
\begin{equation}\label{Hadiabatic}
\mathcal{H}^{adia}_{{SCIM}}(c_\#)
=
V_{{SCIM}}(c_\#)
+
\left[
B_{{SCIM}}(c_\#)
\frac{\partial}{\partial c_\#}
\right]^{(2)}.
\end{equation}
where the dissipative tensor $D_{SCIM}$ disappears. 
As discussed previously, the numerical stability of the SCIM construction critically depends on the regularity of the kernel moments. Despite the introduction of the overlap-based collective coordinate $c_\#$, we found that the kernel moments may still exhibit rapid local variations, preventing a reliable evaluation of the SCIM Hamiltonian.

\begin{figure}
\centering
\includegraphics[width=1.0\linewidth]{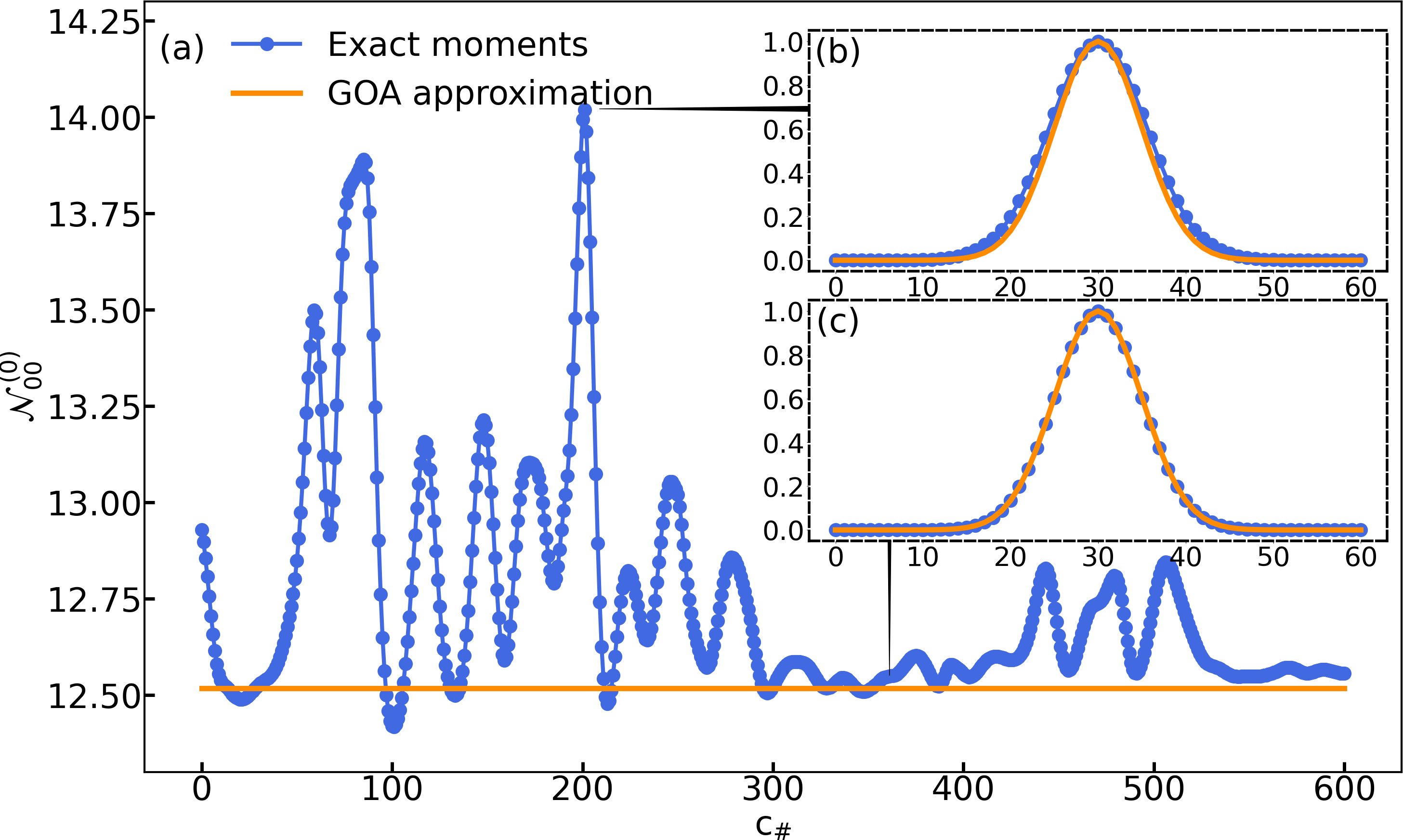}
\caption{Panel (a): Exact and GOA moments $\mathcal{\bar N}_{00}^{(0)}$ with respect to the new collective variable $c_\#$. Panel (b): Exact kernels $\bra{\Phi(\bar q - s)}\ket{\Phi(\bar q + s)}$ for $\bar q$ = 201, compared with the GOA kernels. Panel (c): Exact kernels $\bra{\Phi(\bar q - s)}\ket{\Phi(\bar q + s)}$ for $\bar q$ = 330, compared with the GOA kernels. Calculations have been performed for the $\mathcal{\tilde P}_{20}$ adiabatic set of the $^{240}$Pu asymmetric path.}
\label{ctwo_51}
\end{figure}

To illustrate this point, we consider the zeroth-order moment of the overlap kernel as a simple indicator of the regularity of the kernel moments:
\begin{eqnarray}
\mathcal{\bar N}_{00}^{(0)}(c_{\#}) = \int ds \bra{\Phi_0(c_{\#}-s)}\ket{\Phi_0(c_{\#}+s)}.
\end{eqnarray}
Within the GOA framework, the overlap kernel is assumed to follow a Gaussian dependence on the relative collective coordinate $s$. As a consequence, the zeroth-order moment of the overlap kernel is expected to remain constant along the collective path, providing a useful reference for assessing deviations from regular behavior. In panel (a) of FIG. \ref{ctwo_51}, we compare the exact moment $\mathcal{\bar N}_{00}^{(0)}$ with the constant GOA value. Panels (b) and (c) show two representative overlap kernels compared with their GOA approximations.

The comparison shows that the GOA provides a remarkably good description of the overlap kernels. The remaining deviations are small and mainly affect the amplitude of the kernels, with a systematic slight underestimation by the GOA. However, these small deviations generate rapid oscillations of the exact moments along the collective coordinate. Although limited in amplitude, their short wavelength is incompatible with the regularity properties required for the construction of the SCIM Hamiltonian.

To regularize the kernel moments, we introduce a Savitzky-Golay (SG) low-pass filter \cite{SGDif}. The SG method replaces local finite-difference derivatives with derivatives of a polynomial least-squares fit constructed over a finite window of size $r$. In this work, we consistently use cubic polynomials ($n=3$). More precisely, at each point $c_\#$, a polynomial is fitted to the kernel moments within the interval
\begin{equation}
\left[
c_\#-\frac{r-1}{2},
\,
c_\#+\frac{r-1}{2}
\right],
\end{equation}

and the derivatives are obtained analytically from the fitted polynomial. Increasing the window size $r$ progressively suppresses shorter-wavelength components while preserving the large-scale behavior of the moments.

\begin{figure}
\includegraphics[width=1.0\linewidth]{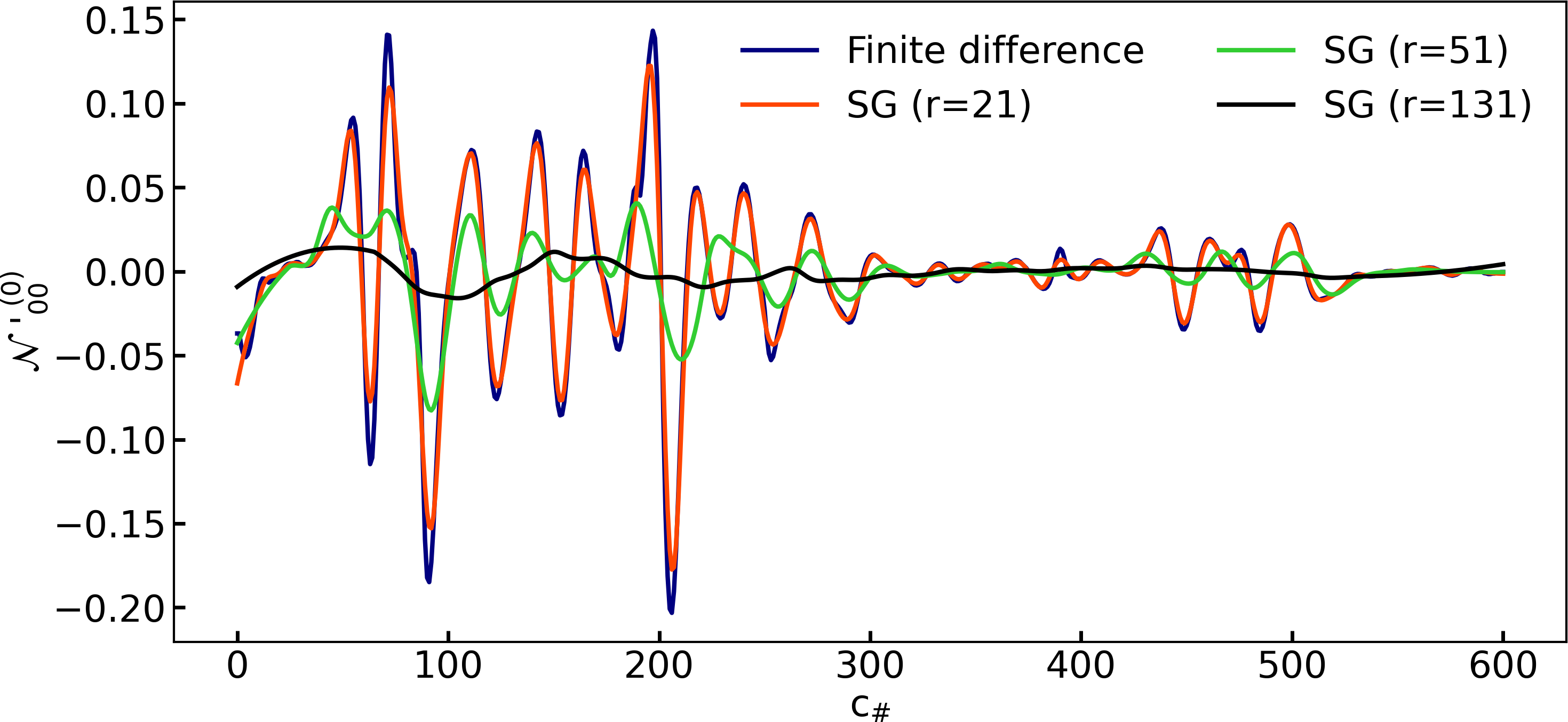}
\caption{Comparison between standard finite differences and SG differentiation in the calculation of the derivative of $\mathcal{N}_{00}^{(0)}$ for different values of the SG parameter $r$.}
\label{cfin_1}
\end{figure}

The need for this regularization is illustrated in FIG. \ref{cfin_1}, where the derivative $\mathcal{N'}_{00}^{(0)}$ of the diagonal overlap moment $\mathcal{N}_{00}^{(0)}$ is evaluated using both standard finite differences and SG differentiation for different values of $r$. The finite-difference derivative exhibits strong high-frequency oscillations, whereas the SG differentiation progressively reduces these components as the window size $r$ is increased. In practice, values of $r\geq25$ are required to obtain a stable evaluation of the inverse square-root operator $\bar{\mathcal N}^{-1/2}$.

\begin{figure}
\includegraphics[width=1.0\linewidth]{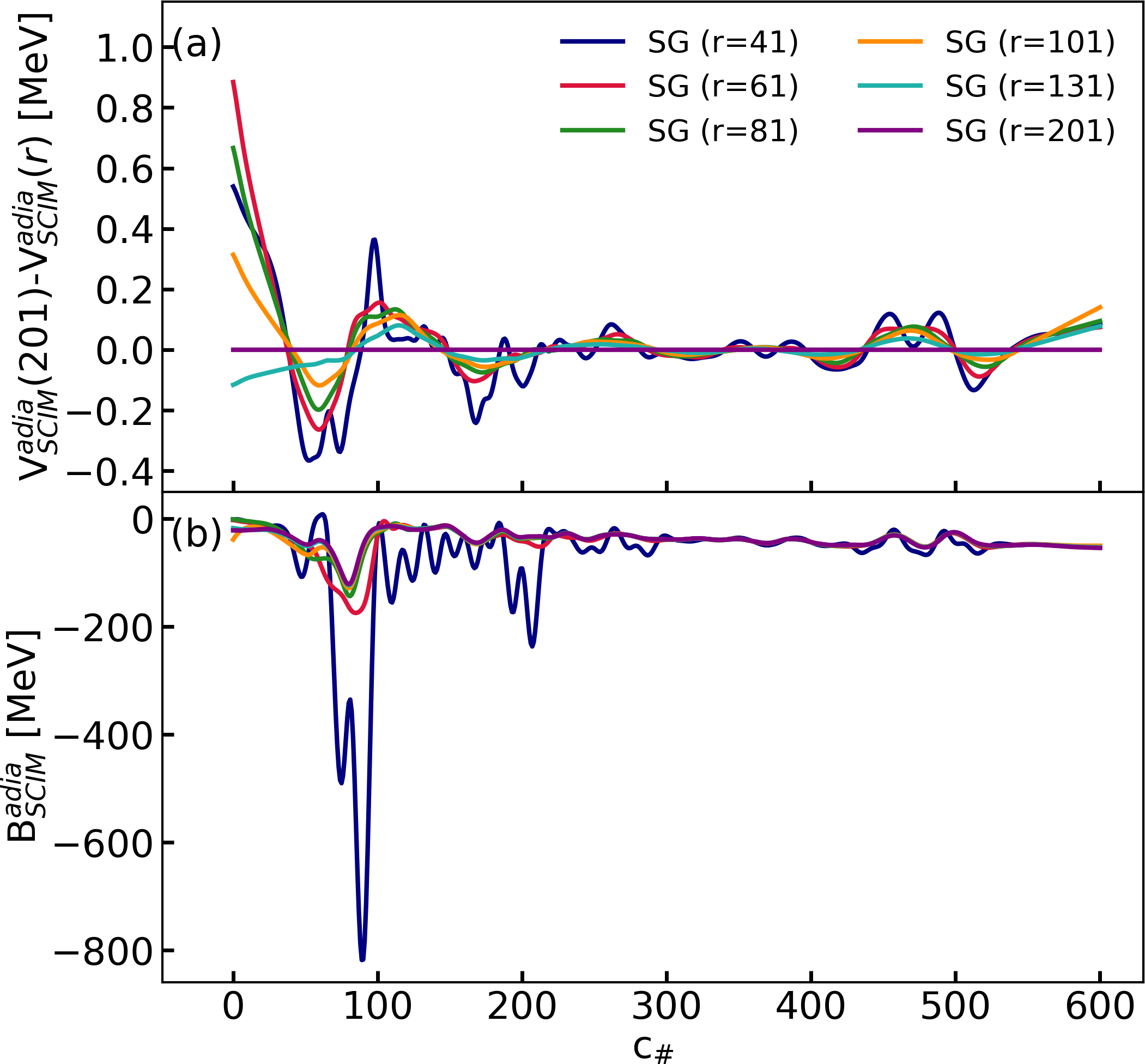}
\caption{Panel (a): Impact of the SG parameter $r$ on the collective potential $V_{{SCIM}}(c_\#)$. The difference $V_{SCIM}^{adia}(r=201) - V_{SCIM}^{adia}(r)$ has been plotted. Panel (b): same for SCIM inertia tensor $B_{{SCIM}}$.}
\label{cfin_2}
\end{figure}

Once the inverse square root operator is stabilized, the collective quantities $V_{{SCIM}}$ and $B_{{SCIM}}$ can be evaluated. Their dependencies on the SG parameter $r$ are shown in FIG. \ref{cfin_2} according to $c_\#$. In the case of $V_{{SCIM}}$, we have represented the difference $V_{SCIM}^{adia}(r=201) - V_{SCIM}^{adia}(r)$.
A striking feature is that the collective potential $V_{{SCIM}}$ remains almost unaffected by the choice of $r$, whereas the inertia tensor $B_{{SCIM}}$ is highly sensitive to the filtering scale. This behavior can be readily explained: the leading contribution to the potential primarily involves non-derivative kernel ratios, which remain unchanged by the SG differentiation procedure. In contrast, the inertia tensor is dominated by derivative contributions and is thus far more sensitive to high-frequency fluctuations.
Moreover, as $r$ increases, the inertia tensor gradually converges toward a stable value. In this work, we have adopted $r=131$ for all applications. This choice ensures satisfactory convergence of the collective inertia while leaving the collective potential essentially unaltered.

The physical motivation behind the SG regularization is closely tied to the structure of the GOA, as discussed below. Specifically, the GOA relies on a second-order expansion of the reduced kernel $h(\alpha',\alpha)$ (see Eq.\ref{reducedh}) around its diagonal $\alpha'=\alpha$, thereby neglecting higher-order non-local variations. In this sense, the GOA inherently acts as a low-frequency approximation of the collective dynamics. The SG regularization serves a closely analogous purpose within the SCIM framework. By filtering out rapid local oscillations in the kernel moments when evaluating their derivatives, it suppresses short-wavelength structures that cannot be consistently described within the second-order truncation underlying the SCIM formulation. Consequently, the SG procedure should not be viewed as a purely numerical smoothing technique. Instead, it represents a controlled regularization consistent with the effective low-frequency character shared by both the GOA and the SCIM formalism.

\subsection{Formal correspondence between the GOA and the SCIM approaches}

In order to perform an analytical comparison between the GOA and the SCIM in the adiabatic limit, we first briefly recall the main features of the GOA. This approximation provides a standard framework to reformulate the Hill-Wheeler equation as a local eigenvalue problem by approximating the overlap kernel $\mathcal{N}(q,q')$ with a Gaussian form in a suitably chosen collective coordinate $q$:

\begin{equation}
\mathcal{N}(q,q') \simeq \exp\left[-\frac{1}{2}\gamma(q) \left(q-q'\right)^2\right].
\end{equation}
In practical applications, the width $\gamma(q)$ is commonly assumed to be constant. However, microscopic calculations indicate a residual dependence on the coordinate. This limitation can be partially alleviated by defining a rescaled coordinate $\alpha(q)$ as:
\begin{eqnarray}
 \alpha(q) = \bigint^q  \sqrt{\frac{\gamma(q')}{\gamma_0}} \,dq' ,
\end{eqnarray}
which ensures, at leading order, a nearly constant Gaussian width $\gamma_0$.

Starting from the Gaussian ansatz for the overlap kernel, the GOA further assumes that the reduced Hamiltonian kernel $h(\alpha',\alpha)$ defined by
\begin{equation}\label{reducedh}
h(\alpha',\alpha)=\frac{\mathcal H(\alpha',\alpha)}{\mathcal N(\alpha',\alpha)}
\end{equation}
can be expanded up to second order around a local collective coordinate $z$. 
By adopting a Gaussian form for the overlap kernel, the powers of $(\alpha-z)$  and $(\alpha'-z)$ can be rewritten as derivative operators acting on the square root of the overlap. Following integration by parts, the Hill-Wheeler equation is reduced to a local collective Schrödinger-like equation, involving solely second-order derivatives with respect to the collective coordinate.
Within this construction, the collective potential and inertia are directly determined by the local derivatives of the kernel $h(\alpha',\alpha)$ evaluated on the diagonal. The GOA thus amounts to a low-frequency (or long-wavelength) approximation of collective dynamics, in which only smooth local variations of the kernels are retained while higher-order non-local structures are neglected. 

Explicitly, the resulting GOA collective Hamiltonian can be written as:
\begin{equation}
\mathcal{H}_{{GOA}} = V_{{GOA}}(z) + \frac{\partial}{\partial z} B_{{GOA}}(z)\frac{\partial}{\partial z},
\end{equation}
where the quantities $V_{{GOA}}$ and $B_{{GOA}}$ are the collective potential and inertia, respectively:
\begin{eqnarray}\label{VBM}\begin{cases}
\displaystyle V_{GOA}(z) = h(z,z) + \frac{1}{2\gamma_0}h^{(2,0)}(z,z) \\ \displaystyle \qquad \qquad \qquad \quad \quad + \frac{1}{8\gamma_0^2}\frac{\partial^2}{\partial z^2}\left[h^{(2,0)}(z,z)\right] .
\\ \displaystyle B_{GOA}(z) = \frac{1}{4\gamma^2_0}\left[ h^{(2,0)}(z,z) - h^{(1,1)}(z,z)\right] \end{cases}
\end{eqnarray}
Here, the quantities $h^{(i,j)}(x,y)$ define multiple partial derivatives such that:
\begin{equation}
h^{(i,j)}(x,y) = \frac{\partial^i}{\partial x^{i}}\frac{\partial^j}{\partial y^{j}} h(x,y).
\end{equation}
In practical TDGCM+GOA calculations, these exact expressions are rarely evaluated directly. Indeed, discontinuities in the manifold of constrained HFB states generally preclude stable numerical evaluation of the required derivatives. For this reason, most applications rely instead on cranking or ATDHFB approximations based on the linear-response theory, which circumvent the explicit calculation of the non-local derivatives entering the exact GOA expressions \cite{walid}. 

Returning to the SCIM formalism in the adiabatic limit, the Hamiltonian given by Eq.~(\ref{Hadiabatic}) can be rewritten using the SOPO definitions. This leads to the following expression:
\begin{eqnarray}\label{cfin_27}
\mathcal{H}^{adia}_{SCIM}(c_\#) = \left[V_{SCIM}(c_\#) + B^{''}_{SCIM}(c_\#)\right] \nonumber\\ + 4\frac{\partial}{\partial c_\#}B_{SCIM}(c_\#)\frac{\partial}{\partial c_\#}. 
\end{eqnarray}
From this equation, a comparison can be perform between the 
GOA and SCIM approach in the adiabatic limit.
As exposed in the following, the SCIM quantities obtained with the SG regularization reproduce remarkably well the GOA masses and zero-point energies evaluated directly from the microscopic kernels. This agreement strongly supports the interpretation of the SG filter as a physically meaningful low-frequency regularization.

\subsection{Comparison of collective dynamical ingredients in SCIM and GOA-based approaches}

In order to assess the quality of the different approximations entering the construction of collective Hamiltonians, we now compare the dynamical ingredients predicted by the SCIM formalism with those obtained within the GOA and the related GOA+Cranking and GOA+ATDHFB local approximations. 

\subsubsection{Zero-point energies}
 
 Rather than focusing on the collective potentials themselves, we consider the associated zero-point energies (ZPE), which provide a more sensitive probe of the underlying collective dynamics. The ZPE associated with a given collective potential $V$ can be defined as:
\begin{equation}
{ZPE}(c_\#)=V(c_\#)-E_{{HFB}}(c_\#),
\end{equation}
where $E_{{HFB}}$ corresponds to the total HFB binding energy. In the case of the GOA approximation, $V(c_\#) \equiv V_{GOA}(c_\#)$ whereas for the SCIM, $V(c_\#) \equiv V_{SCIM}(c_\#) + B^{''}_{SCIM}(c_\#)$.

\begin{figure}
\centering
\includegraphics[width=1.0\linewidth]{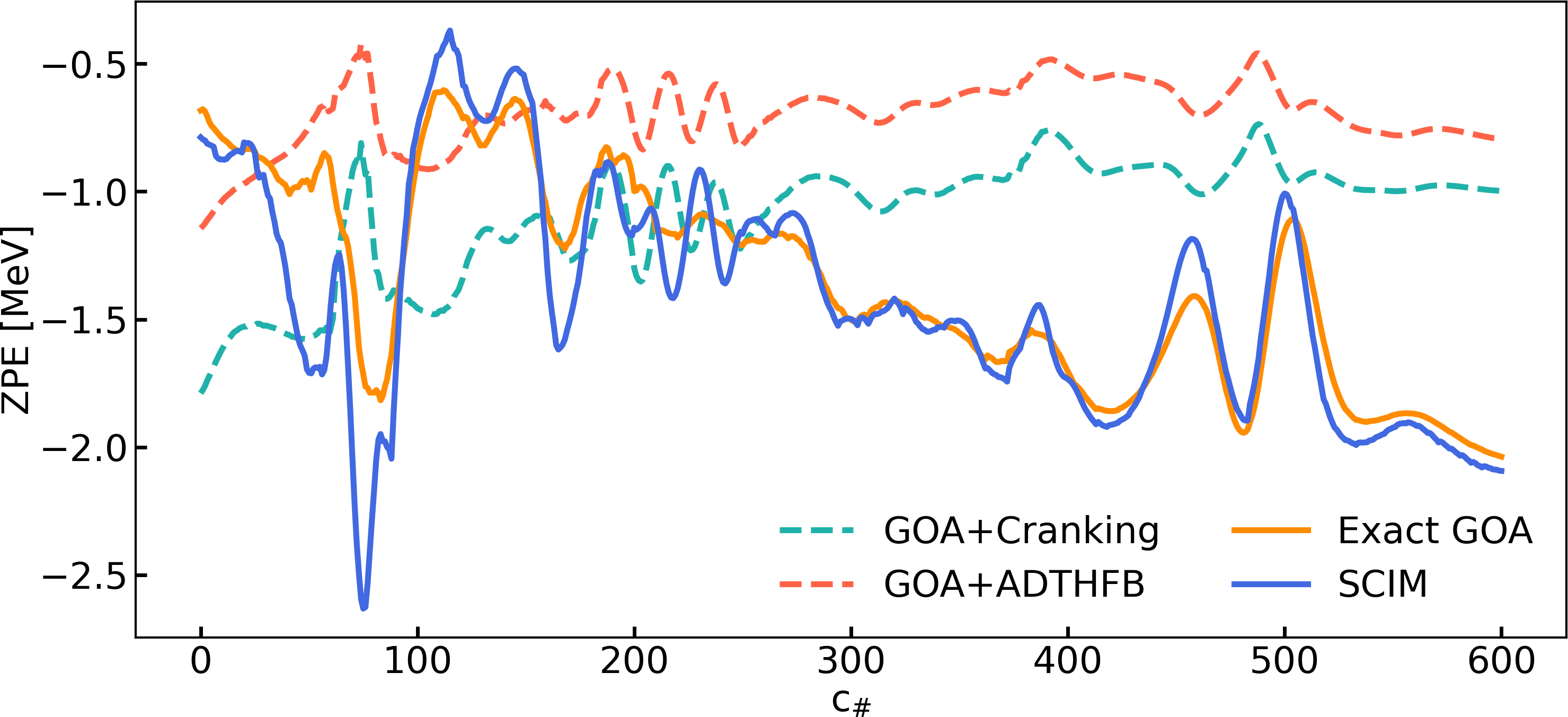}
\caption{Comparison between the ZPE obtained within the SCIM formalism, the exact GOA expressions, the GOA+Cranking and the GOA+ATDHFB approximations.}
\label{cfin_4}
\end{figure}

In FIG. \ref{cfin_4}, the ZPEs evaluated directly from the microscopic kernels of the SCIM and exact GOA approaches are displayed along the asymmetric fission path of $^{240}$Pu. The ones obtained from the standard GOA+Cranking and GOA+ATDHFB approximations are also shown. The path is generated using the $\mathcal{\tilde P}_{20}$ procedure combined with the \enquote{Link} and \enquote{Drop} protocols.

The most remarkable result is the near-perfect agreement between the SCIM and exact GOA predictions. This agreement is non-trivial, as both approaches rely on conceptually distinct constructions of the collective Hamiltonian. It demonstrates that the SG-regularized SCIM framework effectively captures the same essential content of the collective dynamics as the second-order GOA expansion, despite their different formulations. Small deviations do, however, appear in the ground-state region ($c_\# \approx 80$) and near the first barrier ($c_\# \approx 120$). Although these differences are moderate, they may have non-negligible implications for tunneling properties and barrier penetration estimates.

The behavior of the GOA+Cranking and GOA+ATDHFB approximations differs qualitatively. Both approaches exhibit a noticeably shifted evolution along the descent toward scission, together with a slight displacement along the collective coordinate. This pattern suggests that these approximations do not merely introduce quantitative variations but also alter the effective representation of the collective path itself.

\subsubsection{Inertial masses}

A similar analysis is performed for the collective masses $M$, which provide a more direct characterization of the collective dynamics. In the SCIM framework, the collective mass is related to the inertia tensor through
\begin{equation}
M_{{SCIM}}(c_\#)=-\frac{1}{2B_{{SCIM}}(c_\#)}.
\end{equation}

\begin{figure}
\centering
\includegraphics[width=1.0\linewidth]{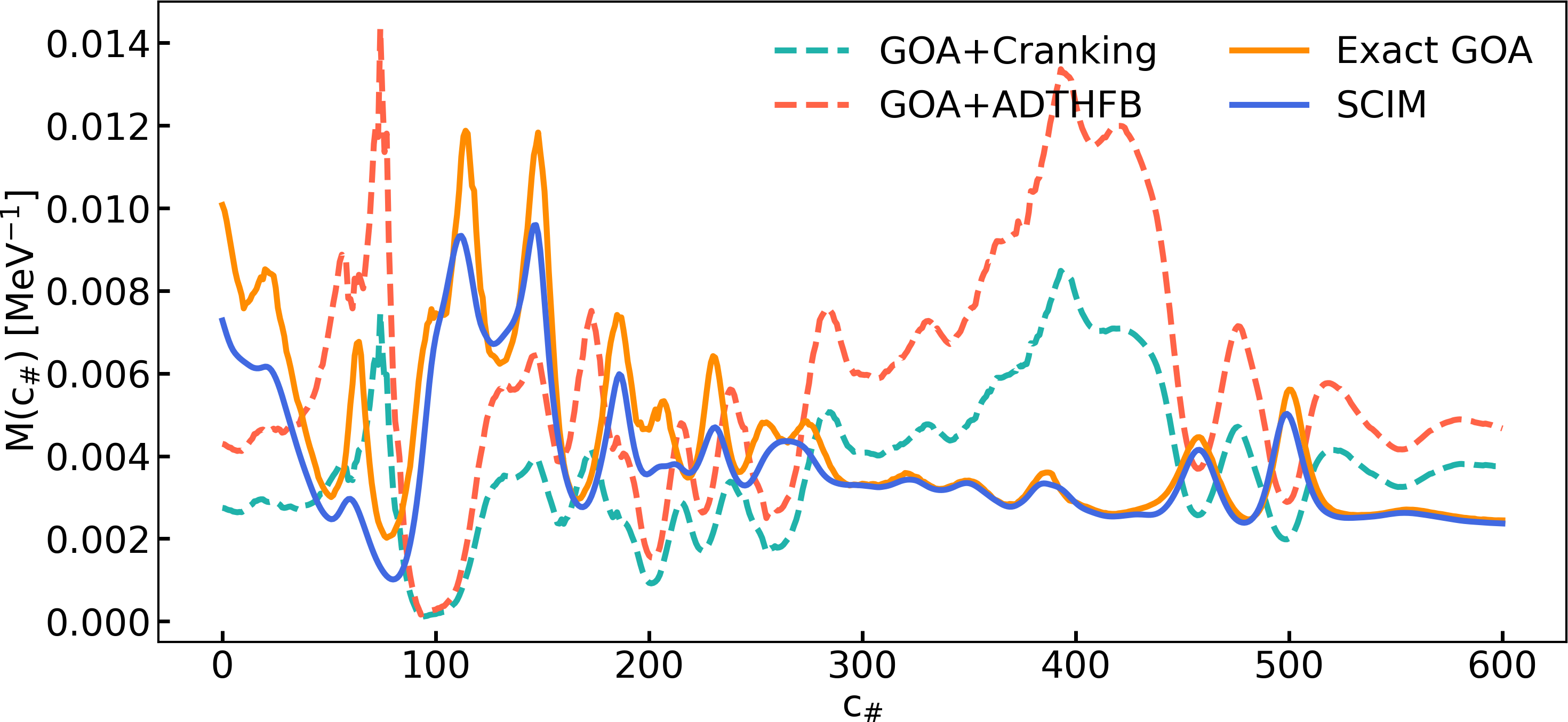}
\caption{Same as FIG. \ref{cfin_4} but for the inertial mass.}
\label{cfin_5}
\end{figure}

FIG. \ref{cfin_5} shows the comparison between SCIM, exact GOA, GOA+Cranking, and GOA+ATDHFB results.
Once again, the agreement between SCIM and exact GOA is striking. This further supports the interpretation that both approaches consistently capture the same essential content of the collective dynamics, despite encoding this information differently in each formalism.

The comparison with the cranking and ATDHFB masses is particularly revealing near the first barrier ($c_\# \approx 120$), where both approximations significantly underestimate the inertial masses and consequently tend to underestimate spontaneous fission half-lives. 
This observation is especially noteworthy, as the underestimation of spontaneous fission half-lives has historically been one of the primary reasons for favoring ATDHFB prescriptions over cranking masses.

Beyond the quantitative differences observed near the first barrier, the results reveal clear qualitative differences between the inertial masses obtained within the exact GOA and SCIM frameworks and those derived from the Cranking and ATDHFB approximations. In particular, along the descent from the saddle point to scission, the inertial masses predicted by the exact GOA and SCIM approaches are significantly smaller than those obtained with the Cranking and ATDHFB approximations. This behavior indicates that the treatment of non-local effects goes beyond a simple quantitative correction of the collective inertia: it leads to a qualitatively different description of the collective dynamics along the fission path.

Taken together, these observations suggest that efforts to improve collective inertial masses should first focus on the treatment of non-locality. Indeed, the present results indicate that the neglect of non-local effects represents a major limitation of both the 
GOA+Cranking and GOA+ATDHFB approaches, not only quantitatively, but also qualitatively, as it leads to a substantially modified dynamical evolution along the 
fission path.
This interpretation should not be seen as dismissing the importance of time-odd effects, which are genuinely expected in microscopic time-dependent theories. 
Nevertheless, the present comparison suggests that part of the success of ATDHFB-based prescriptions may arise from an effective compensation for missing 
non-local collective correlations. 

\section{Excited collective dynamical ingredients}\label{dynaing}

In this section, we analyze the diagonal and off-diagonal properties of the collective potential $V_{SCIM}$, inertia tensor $B_{SCIM}$, and dissipative tensor $D_{SCIM}$ when intrinsic excitations are included in the SCIM framework. This analysis provides insight into the mechanisms underlying the emergence of dissipative effects.

The SG regularization introduced in Sec.~\ref{savgol} is applied to the three quantities introduced above. In the following, we refer to the \enquote{adiabatic-excited} limit as the case where each variational excitation is considered independently, without coupling between different excited configurations. More explicitly, each excited set is analyzed using the same procedure as for the adiabatic set.

\subsection{Collective potential $V_{SCIM}$}

We first discuss the SCIM collective potential in the \enquote{adiabatic limit} and for the six variational excitations considered in the dynamical calculations within the \enquote{adiabatic-excited} limit.

\begin{figure}
\centering
\includegraphics[width=1.0\linewidth]{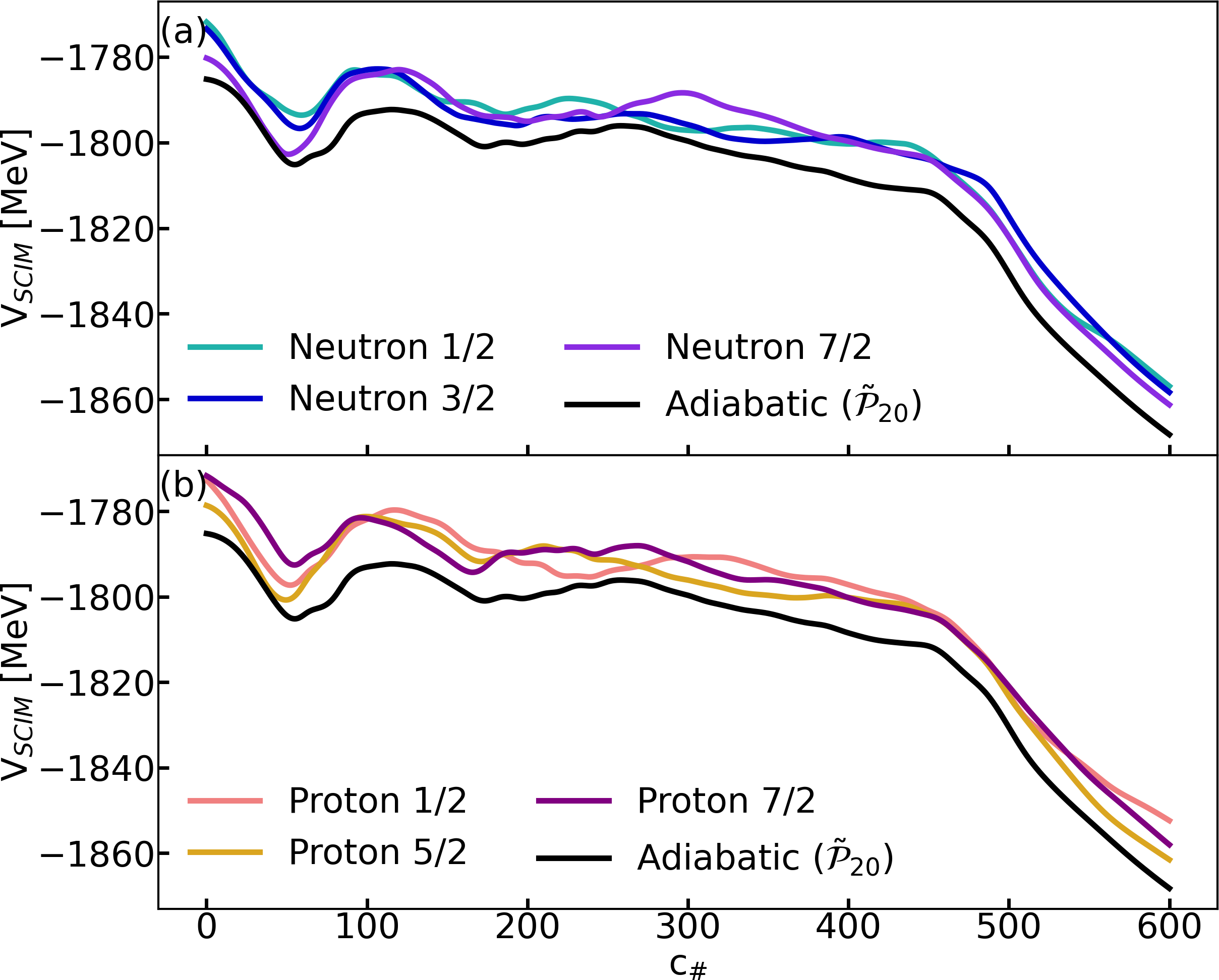}
\caption{Panel (a): SCIM collective potential at the adiabatic limit (in black) and SCIM potentials at the \enquote{adiabatic-excited} limit associated with the neutron variational excitations. Panel (b): same as panel (a) but for proton variational excitations. }
\label{cfin_6}
\end{figure}

In detail, panel (a) of FIG. \ref{cfin_6} shows the three neutron excitations with $\Omega$ equal to $1/2$, $3/2$, and $7/2$, built on top of the adiabatic asymmetric path of $^{240}$Pu and discussed in the second article of the trilogy. Panel (b) shows the corresponding three proton excitations with $\Omega$ equal to $1/2$, $5/2$, and $7/2$. The adiabatic set is represented by black solid lines. All curves are plotted as functions of the collective coordinate $c_\#$.

No irregular behavior is observed for the SCIM potentials shown in either panel. The differences between the various potentials remain of the same order of magnitude as the corresponding differences between their HFB energies (see the second article of this trilogy for comparison).\\

\begin{figure}
\centering
\includegraphics[width=1.0\linewidth]{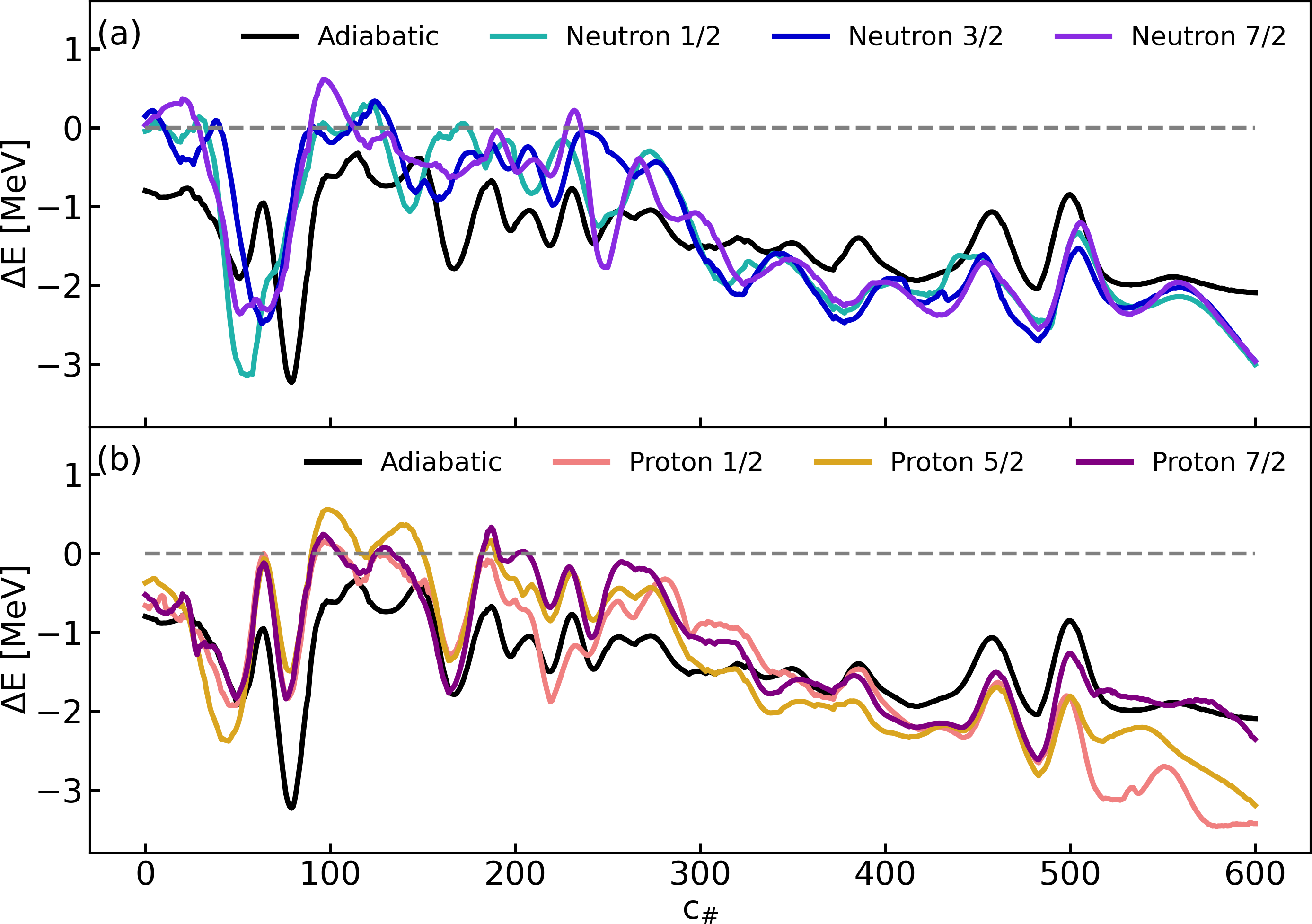}
\caption{Comparison of the energy difference $\Delta E^{ii}$ obtained for the adiabatic state and excitations at the SCIM level. Each curve corresponds to one $\Omega$ for neutron (panel (a)) and for proton (panel (b)).}
\label{cfin_6_bis}
\end{figure}

In FIG. \ref{cfin_6_bis}, we display the evolution of the energy difference $\Delta E^{ii}$ between the diagonal contribution $V_{SCIM}^{ii}$ of the SCIM potential and the corresponding HFB total binding energy $E_{HFB}^{ii}$ for each excitation $i$, as a function of the collective coordinate $c_\#$. This quantity is defined as
\begin{eqnarray}
\Delta E^{ii}(c_\#) = V_{SCIM}^{ii}(c_\#) - E_{HFB}^{ii}(c_\#).
\end{eqnarray}
Panels (a) and (b) correspond to neutron and proton excitations, respectively. The corresponding quantity for the adiabatic state is also shown in black.

All $\Delta E^{ii}$ exhibit a similar behavior and remain of the same order of magnitude. Before the saddle point, the adiabatic $\Delta E^{ii}$ is slightly larger than the values associated with the excited states, whereas the opposite trend is observed beyond the saddle point. In the second well region, the excited-state values of $\Delta E^{ii}$ occasionally reach small positive values, limited to a few hundred keV.

\begin{figure}
\centering
\includegraphics[width=1.0\linewidth]{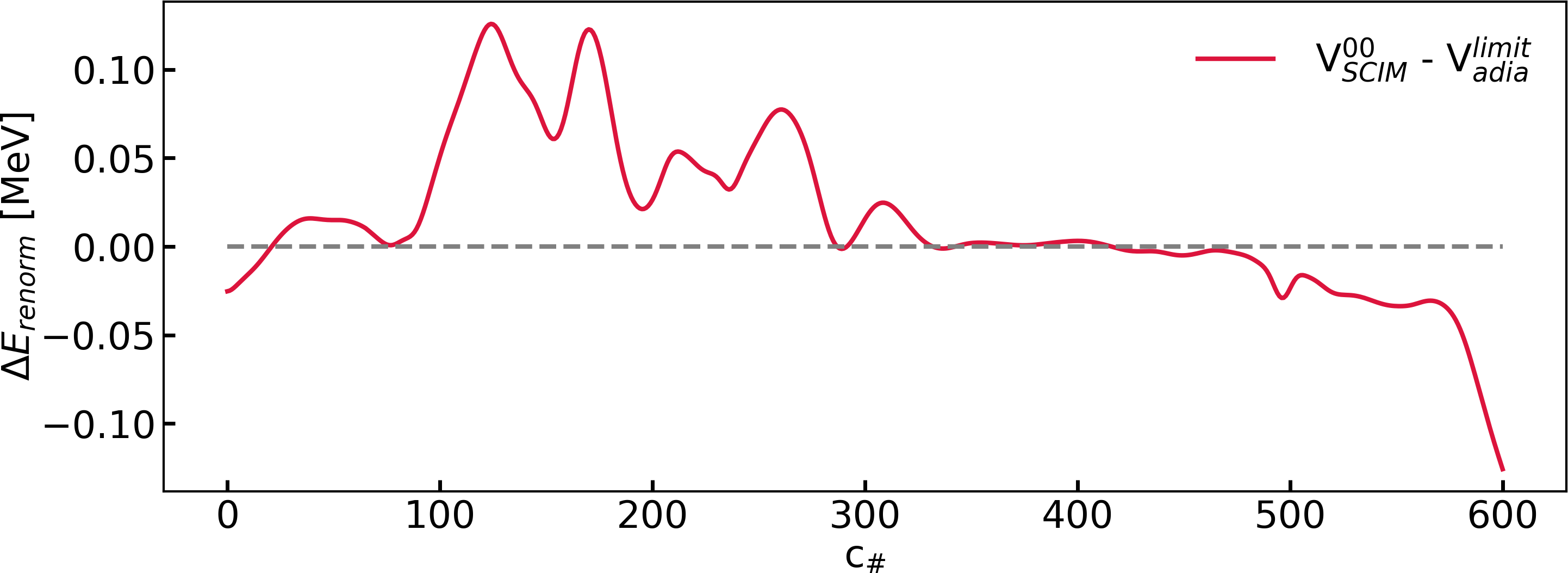}
\caption{Difference $\Delta E_{renorm}$ (in MeV) between the diagonal value V$^{00}_{SCIM}$ of the SCIM potential with and without excited states, according to the collective coordinate $c_\#$.}
\label{Vrenorm}
\end{figure}

To conclude the discussion of the diagonal contribution of the SCIM potential, we investigate the renormalization of the potential induced by the inclusion of excitations. For this purpose, we compare the adiabatic-limit potential $V^{limit}_{adia}$ with the diagonal component $V^{00}_{SCIM}$ obtained when the six excitations are included:
\begin{equation}
\Delta E_{renorm}(c_\#)=V^{00}_{SCIM}(c_\#)-V^{limit}_{adia}(c_\#).
\end{equation}
The corresponding evolution is shown in FIG. \ref{Vrenorm}. The renormalization remains very small with respect to the diagonal $V^{limit}_{adia}$, with a maximum amplitude of about $0.1$ MeV. The small fluctuations mainly follow the underlying structural effects and are essentially located in the second-well region, where $\Delta E_{renorm}$ becomes slightly positive. This behavior is consistent with the trends observed previously for the excited-state quantities $\Delta E^{ii}$. \\

\begin{figure}
\centering
\includegraphics[width=1.0\linewidth]{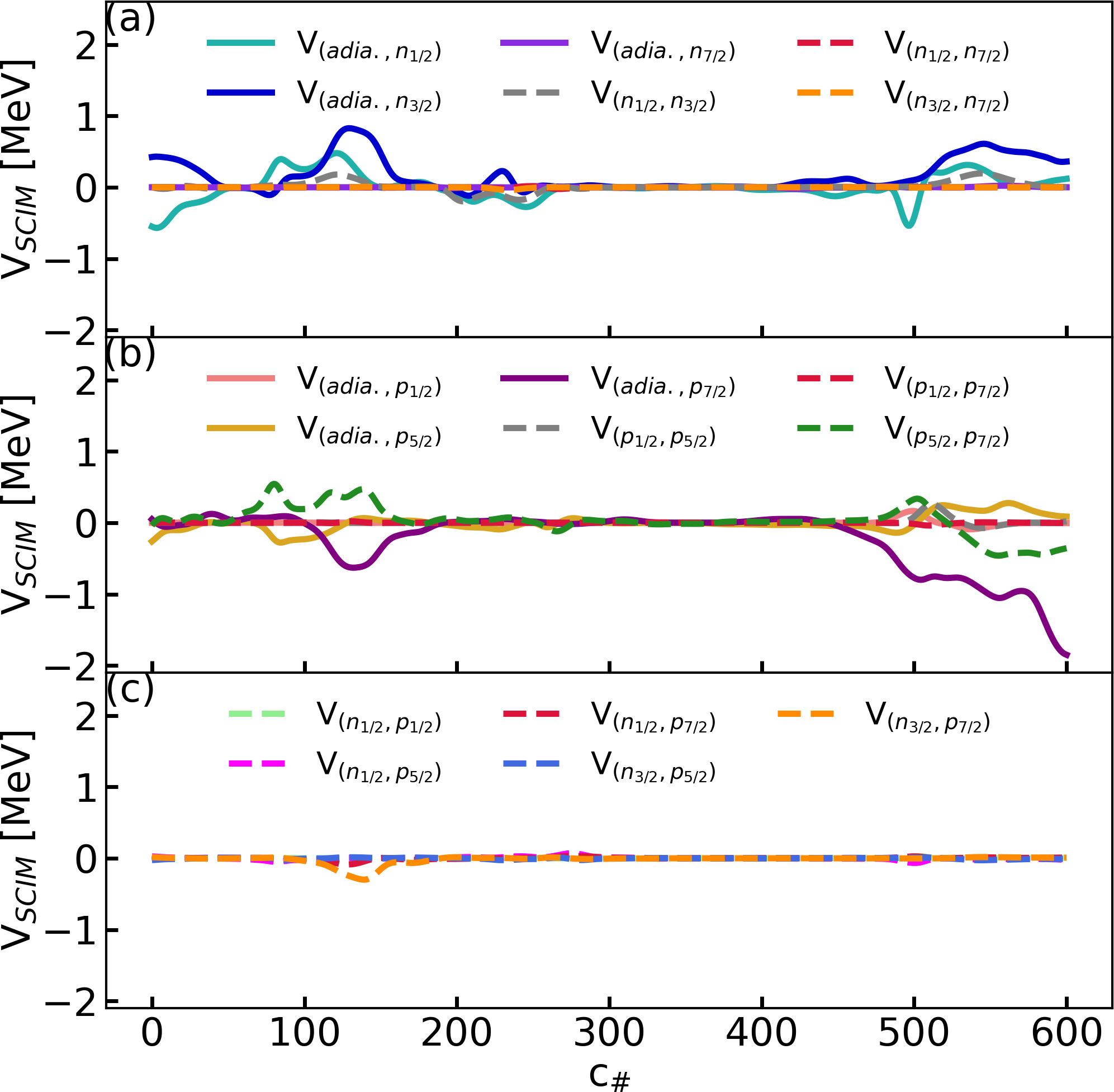}
\caption{Dominant off-diagonal potential components $V^{ij}_{SCIM}$ with respect to $c_\#$. Panel (a): neutron couplings. Panel (b): proton couplings. Panel (c): neutron-proton couplings.}
\label{cfin_8}
\end{figure}

We finally analyze the off-diagonal components of the SCIM collective potential $V_{SCIM}$. FIG. \ref{cfin_8} displays the dominant coupling terms, i.e., those exhibiting non-negligible amplitudes along at least part of the collective path. Panels (a), (b), and (c) correspond to neutron-neutron, proton-proton, and neutron-proton couplings, respectively.

The largest couplings are found in the neutron and proton sectors and involve the adiabatic configuration, mainly in the second-well and scission regions. In contrast, neutron-proton couplings remain negligible. The maximum coupling reaches $1.85$~MeV for the adiabatic-proton $7/2$ proton coupling, while the largest neutron contribution is $0.83$~MeV for the adiabatic-$n_{3/2}$ coupling.

\subsection{Collective inertia tensor $B_{SCIM}$}

\begin{figure}
\includegraphics[width=1.0\linewidth]{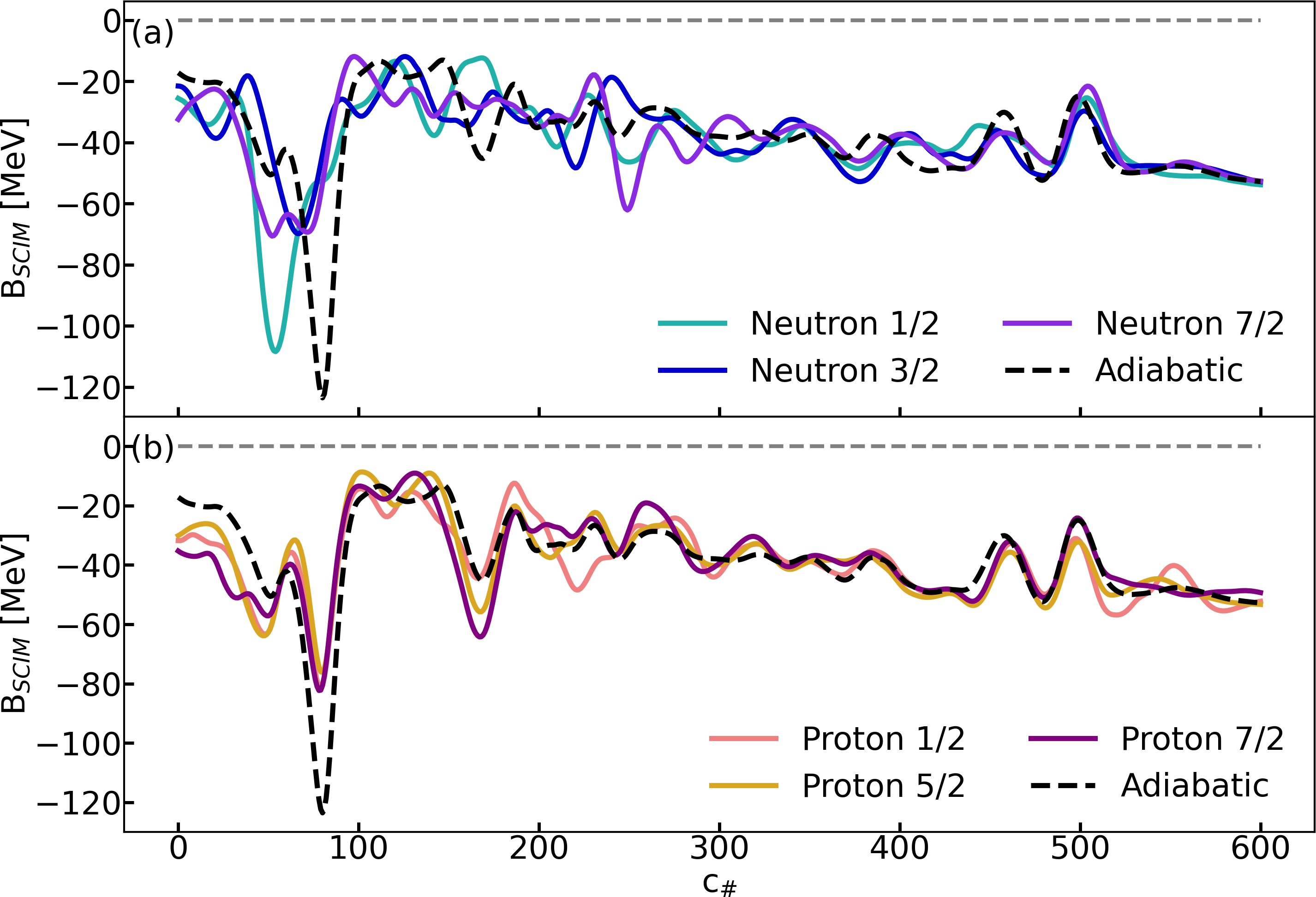}
\caption{Adiabatic-excited inertial tensor $B_{SCIM}$ related to the neutron (panel (a)) and proton (panel (b)) variational excitations used in the present dynamical study. Inertial tensor at the adiabatic limit is also indicated (in black dashed lines).}
\label{cfin_34_bis}
\end{figure}

In this section, we analyze the properties of the collective inertia tensor $B_{SCIM}$. As in the adiabatic case, we first consider the associated collective masses in the \enquote{adiabatic-excited} limit for the neutron and proton excited states. The corresponding quantities are shown in FIG. \ref{cfin_34_bis}, panels (a) and (b), respectively. The adiabatic result is also displayed for comparison (black dashed line).

The overall behavior remains similar to the adiabatic case. For the neutron excitations, the amplitudes remain of the same order of magnitude, with differences mainly appearing through oscillations in the second-well region, reflecting the distinct underlying structures of the excited configurations. For the proton excitations, the $5/2$ and $7/2$ states exhibit larger peaks in this region, with amplitudes approximately $1.8$ times higher than in the other cases.

\begin{figure}
\centering
\includegraphics[width=1.0\linewidth]{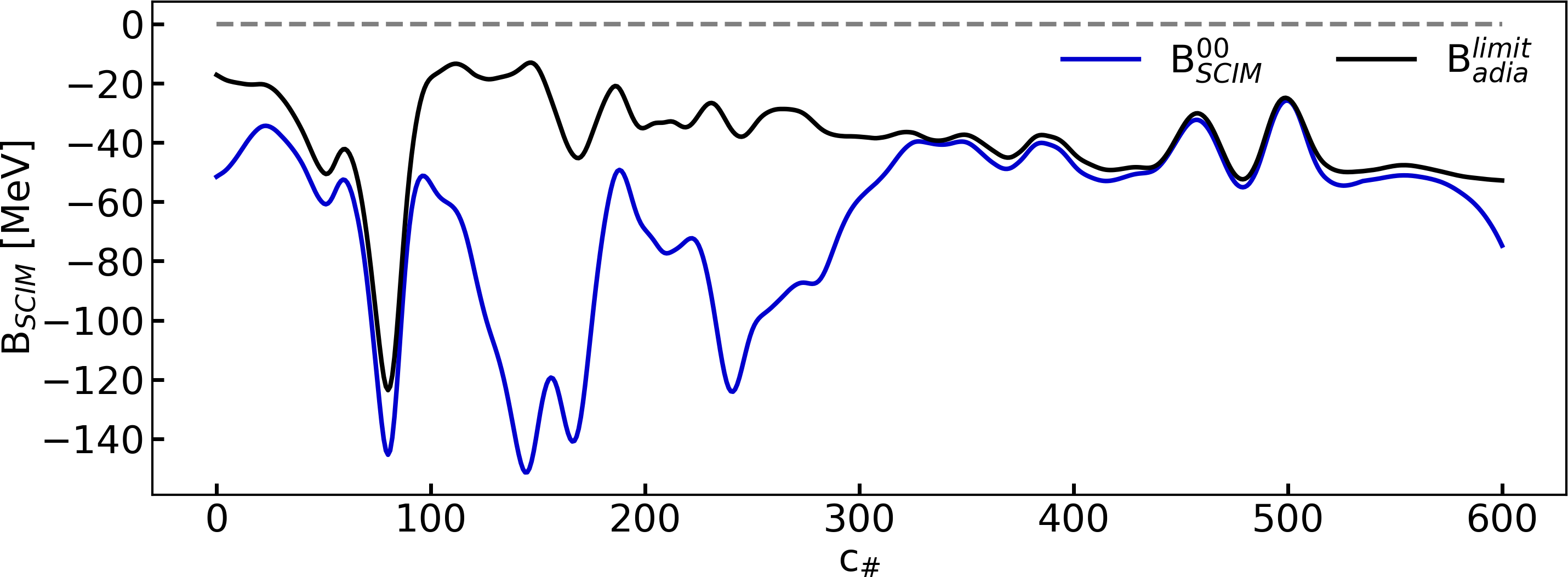}
\caption{Inertia tensor $B^{00}_{SCIM}$ and $B^{limit}_{adia}$ associated with the diagonal value B$_{SCIM}$ of the SCIM inertia tensor $B_{SCIM}$ with and without excited states, respectively.}
\label{Brenorm}
\end{figure}

To conclude our discussion of the diagonal contribution to the SCIM inertia tensor, we examine the renormalization induced by the inclusion of excited configurations. To this end, we compare the inertia tensor obtained in the adiabatic limit, $B^{\mathrm{limit}}_{adia}$, with the diagonal component $B^{00}_{SCIM}$ resulting from the inclusion of the six excited states. The corresponding evolution is shown in FIG.\ref{Brenorm}.
In contrast to the collective potential, the renormalization of the inertia tensor is found to be substantial, reaching amplitudes of about $120$~MeV. This already highlights the profound impact of excited configurations on the collective dynamics. Moreover, the renormalization exhibits two distinct regimes along the collective path: it remains weak in the ground-state well and beyond the saddle point, while becoming much stronger throughout the intermediate first-barrier to saddle region.

\begin{figure}
\centering
\includegraphics[width=1.0\linewidth]{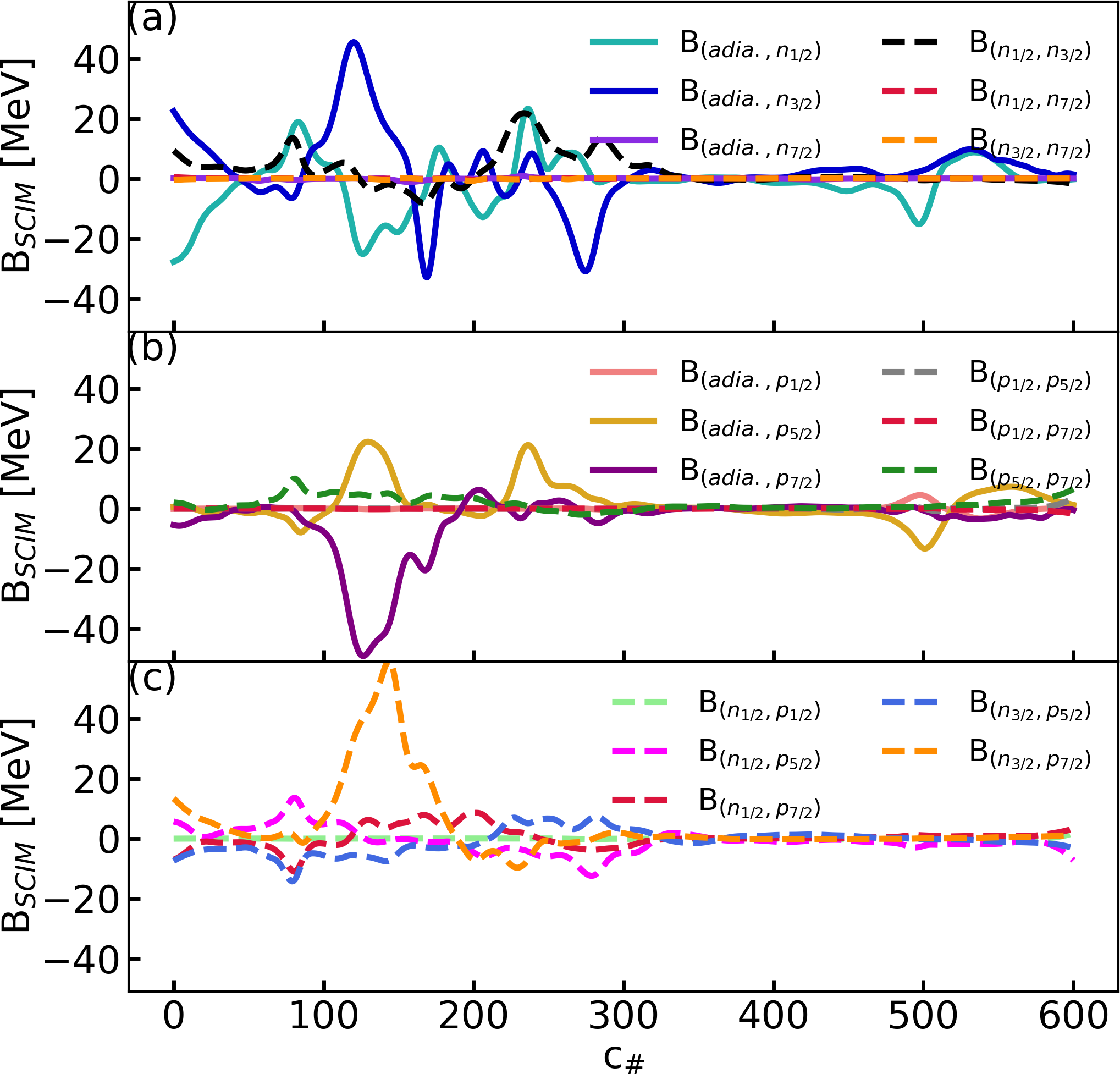}
\caption{Dominant off-diagonal inertia tensor components $B^{ij}_{SCIM}$ with respect to $c_\#$. Panel (a): neutron-neutron couplings. Panel (b): proton-proton couplings. Panel (c): neutron-proton couplings.}
\label{cfin_11}
\end{figure}

The off-diagonal components of the collective inertia tensor are shown in FIG. \ref{cfin_11} as function of the collective coordinate $c_\#$ along the asymmetric fission path of $^{240}$Pu. Their amplitudes become comparable to the diagonal contributions in the second-well and scission regions. However, while the diagonal components remain centered around values of about $-40$~MeV, the off-diagonal terms fluctuate around zero.
Similarly to the collective potential, the dominant couplings involve the adiabatic configuration and the lowest neutron and proton excitations. Around $c_\#\simeq120$, the largest amplitudes are obtained for the adiabatic-proton $7/2$ and adiabatic-neutron $3/2$ couplings, reaching absolute values of $49.15$~MeV and $45.53$~MeV, respectively. The adiabatic-neutron $1/2$ and adiabatic-proton $5/2$ couplings reach $27.72$~MeV and $22.28$~MeV.

Additional couplings appear along the collective path, in particular near scission. The largest off-diagonal contribution is obtained for the coupling between the neutron $3/2$ and proton $7/2$, with an amplitude of $59.02$~MeV around $c_\#\simeq150$. A non-negligible coupling between the neutron $1/2$ and $3/2$ excitations is also observed around $c_\#\simeq250$. Off-diagonal couplings vanish along the descent from the saddle point toward scission.

\subsection{Collective dissipative tensor $D_{SCIM}$}

The collective dissipative tensor $D_{SCIM}$ is a specific feature of the SCIM framework arising from the coupling with excited configurations. It depends on the collective velocity and contributes to the coupling between the different collective potential energy paths.

\begin{figure}
\centering
\includegraphics[width=1.0\linewidth]{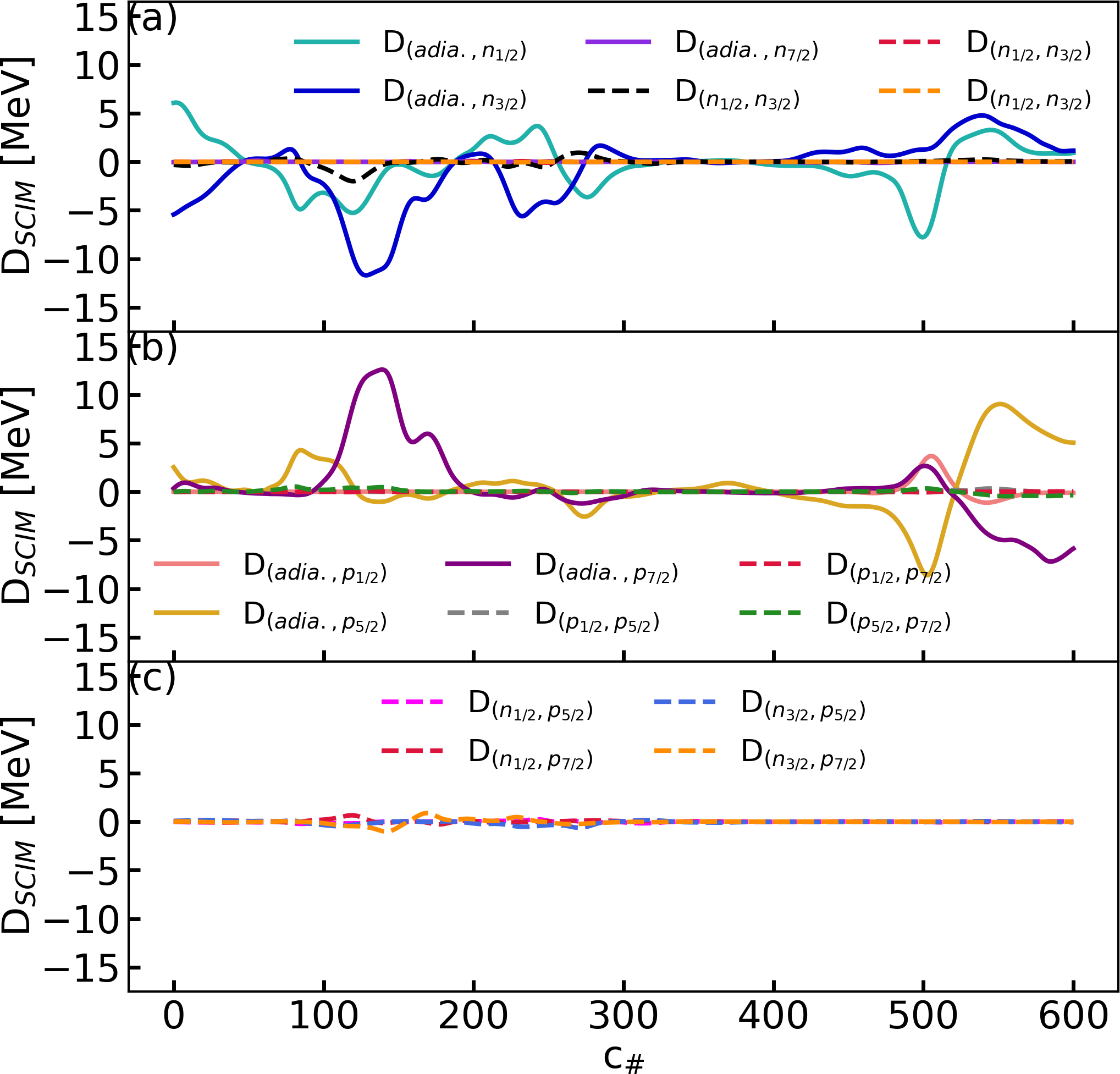}
\caption{Dominant off-diagonal dissipative tensor components $D^{ij}_{SCIM}$ with respect to $c_\#$. Panel (a): neutron couplings. Panel (b): proton couplings. Panel (c): neutron-proton couplings.}
\label{cfin_12}
\end{figure}

Because $D_{SCIM}$ is an antisymmetric operator, its diagonal components vanish. FIG. \ref{cfin_12} displays the dominant off-diagonal components along the asymmetric one-dimensional path of $^{240}$Pu. 
The neutron-proton contributions are found to be much smaller than the pure neutron and proton contributions, and play only a minor role in the present study. As for the collective potential and inertia tensor, the dominant couplings involve the neutron and proton excitations already identified above, mainly in the second-well and scission regions.

Along the descent from the saddle point toward scission, the $D_{SCIM}$ couplings become negligible. This indicates that, once the dissipative coupling is established, its contribution remains weak during the descent phase, whereas significant energy exchanges between the adiabatic and excited configurations are expected near the scission region.

The largest amplitudes are obtained for the adiabatic-neutron $3/2$ and adiabatic-proton $7/2$ couplings, reaching absolute values of $11.70$~MeV and $12.59$~MeV, respectively, around $c_\#\simeq130$. Other non-negligible contributions involve the proton $5/2$, neutron $1/2$, and proton $1/2$ excitations, with maximum amplitudes of $9.04$~MeV, $7.78$~MeV, and $3.68$~MeV, respectively. An excited-excited coupling between the neutron $1/2$ and $3/2$ states is also observed around $c_\#\simeq120$, with an amplitude of about $2$~MeV.

Overall, the off-diagonal contributions of the inertia tensor $B_{SCIM}$ remain larger than those of the dissipative tensor $D_{SCIM}$ in the second-well region, whereas both contributions become comparable near scission. In contrast, the off-diagonal components of the collective potential $V_{SCIM}$ remain significantly smaller along the whole path.

\section{The collective-intrinsic Schr\"odinger equation}\label{propagat}
Once the SCIM Hamiltonian has been constructed \cite{TPaul}, the collective dynamics are obtained by propagating a suitably chosen initial wave packet $g(t)$ according to the Schr\"odinger equation introduced in the first article of this trilogy \cite{trilogy1}. In the following, we refer to this equation as the \enquote{collective-intrinsic Schr\"odinger equation}. It reads
\begin{eqnarray}\label{cfin_37}
 \mathcal{H}_{SCIM} g(t) = i \hbar \frac{\partial}{\partial t}g(t).
\end{eqnarray}
When excited configurations are included, the collective-intrinsic Schr\"odinger equation takes the following matrix form:
\begin{widetext}
\begin{eqnarray}\label{cfin_38}
 \begin{pmatrix}
 (\mathcal{H}_{SCIM})_{00} & \hdots &  (\mathcal{H}_{SCIM})_{0n} \\
 \vdots & \ddots & \vdots \\
(\mathcal{H}_{SCIM})_{n0} & \hdots &  (\mathcal{H}_{SCIM})_{nn}
 \end{pmatrix}
 \begin{pmatrix}
 g_0(t) \\
 \vdots \\
 g_n(t)
 \end{pmatrix}
 = i \hbar \frac{\partial}{\partial t}g(t)
 \begin{pmatrix}
 g_0(t) \\
 \vdots \\
 g_n(t)
 \end{pmatrix}
\end{eqnarray}
\end{widetext}
with $0$ referring to the lowest (adiabatic) state and $n>0$ to the excited states.

In the remainder of this section, we describe the numerical treatment of Eq.~(\ref{cfin_38}). We first present the construction of the initial wave packet and then discuss its time propagation using the Crank-Nicolson scheme \cite{DynaGou,CraNo1,CraNo2}, together with the implementation of the boundary conditions. Finally, we derive the probability fluxes required to extract the fragment yields and excitation properties at scission. To the best of our knowledge, these flux expressions have not been reported previously.

\subsection{Numerical solution}

This section summarizes the numerical implementation of the collective-intrinsic Schr\"odinger equation. We first describe the construction of the initial wave packet and then the numerical procedure used for its time propagation.

\subsubsection{Construction of the initial wave packet}

Two main strategies are commonly used in the literature to construct the initial wave packet. The first consists in building a wave packet confined in an extrapolated ground-state potential well (see, for example, Ref.~\cite{DynaGou}), while the second additionally applies an initial boost along the fission direction (see, for example, Ref.~\cite{DynaRe}). In this first application of the SCIM, we adopt the former approach and restrict the initial state to the adiabatic channel.

The initial wave packet is built from the auxiliary Hamiltonian
\begin{eqnarray}\label{cfin_39}
\mathcal{H}_{ext}(c_\#) = V_{ext}(c_\#) + \left[B_{00}(c_\#)\frac{\partial}{\partial c_\#}\right]^{(2)},
\end{eqnarray}
where $V_{ext}$ is an extrapolated parabolic potential describing the ground-state well and $B_{00}$ denotes the diagonal adiabatic component of the SCIM inertia tensor. Since only the adiabatic channel is considered, no dissipative term appears in Eq.~(\ref{cfin_39}).

Diagonalization of $\mathcal{H}_{ext}$ provides a set of eigenstates $v_i$ that form the basis used to construct the initial wave packet. Their eigenvalues are not used directly because the target energy is defined with respect to the full SCIM Hamiltonian rather than $\mathcal{H}_{ext}$. Instead, each basis state is assigned the expectation value
\begin{eqnarray}
\tilde E_v = \sum_{c_\#} v(c_\#) \mathcal{H}_{SCIM}(c_\#) v(c_\#).
\end{eqnarray}
The initial wave packet is then defined as the Gaussian superposition
\begin{eqnarray}
v(E) =\sum_i v_i \exp\!\left[-\frac{(\tilde E_i-E)^2}{2\sigma_G^2}\right].
\end{eqnarray}
The centroid $E_d$ is determined by dichotomy so that the average energy of $v(E_d)$ equals the prescribed value $\bar E_G$, while its energy standard deviation remains close to $\sigma_G$. The initial wave packet is therefore
\begin{eqnarray}
g(t=0)=v(E_d).
\end{eqnarray}
In the present work, we choose
$\bar E_G=-1792.21$~MeV, corresponding to the top of the first barrier of the adiabatic potential, together with an energy standard deviation of $\sigma_G=0.5$~MeV. These values are representative of the low-energy fission regime targeted by the SCIM framework.

\subsubsection{Time propagation}

The collective-intrinsic Schr\"odinger equation is solved using the Crank-Nicolson scheme, which leads to
\begin{eqnarray}\label{cfin_40}
\frac{g(c_\#,t+ \Delta t) - g(c_\#,t)}{\Delta t} = - \frac{i \mathcal{H}_{SCIM}(c_\#)}{2 \hbar} \qquad \nonumber \\ \times \left[ g(c_\#,t+ \Delta t) + g(c_\#,t) \right]. 
\end{eqnarray}
This expression can be rewritten as the linear system
\begin{eqnarray}
(1 + i \frac{\mathcal{H}_{SCIM}(c_\#)\Delta t}{2 \hbar})g(c_\#,t+ \Delta t) = \qquad \qquad \nonumber\\
(1 - i \frac{\mathcal{H}_{SCIM}(c_\#)\Delta t}{2 \hbar})g(c_\#,t). 
\end{eqnarray}
Following Ref.~\cite{DynaGou}, this system is solved iteratively. Denoting by
$g^{(i)}(t+\Delta t)$
the successive iterates and choosing
$g^{(0)}(t+\Delta t)=g(t)$,
the iteration reads
\begin{eqnarray}\label{cfin_41}
g^{(i+1)}(c_\#,t + \Delta t) = (1 - i \frac{\mathcal{H}_{SCIM}(c_\#)\Delta t}{2 \hbar})g(c_\#,t) \qquad \nonumber \\ - i \frac{\mathcal{H}_{SCIM}(c_\#)\Delta t}{2 \hbar} g^{(i)} (c_\#,t + \Delta t). 
\end{eqnarray}

Throughout this work, a time step
$\Delta t = 6\times10^{-4}\hbar$
is used, which was found to provide stable numerical propagation.

Since the collective coordinate is discretized on a finite interval, spurious reflections naturally occur at the boundaries ($c_\#=0$ and $c_\#=600$). To suppress them, the SCIM Hamiltonian is extended by an absorbing Hamiltonian $\mathcal{H}_{abs}$ acting over the interval $[601,800]$. This additional region progressively absorbs the outgoing wave packet before it reaches the numerical boundary.

To avoid artificial couplings inside the absorbing layer, all off-diagonal components of $V_{abs}$, $D_{abs}$ and $B_{abs}$ are set to zero. The diagonal components of the inertia tensor are kept constant,
\begin{eqnarray}
(B_{abs})_{ii}(c_\#)
=
(B_{SCIM})_{ii}(600),
\end{eqnarray}
whereas the real part of the potential is obtained by linear extrapolation of the SCIM potential. Its imaginary part is chosen as a quadratic absorbing potential whose coefficients were adjusted empirically to minimize spurious reflections. The resulting diagonal potential reads
\begin{eqnarray}
(V_{abs})_{ii}(c_\#) = \qquad \qquad \qquad \qquad \qquad \qquad \qquad \qquad  \nonumber \\ \left(c_\# -600\right) \left[(V_{SCIM})_{ii}(600) -  (V_{SCIM})_{ii}(599)\right] \nonumber \\ + (V_{SCIM})_{ii}(600) - i\left[-5.10^{-6} c_\#^2 + 0.1c_\# -62\right].
\end{eqnarray}

\begin{figure}
\centering
\includegraphics[width=1.0\linewidth]{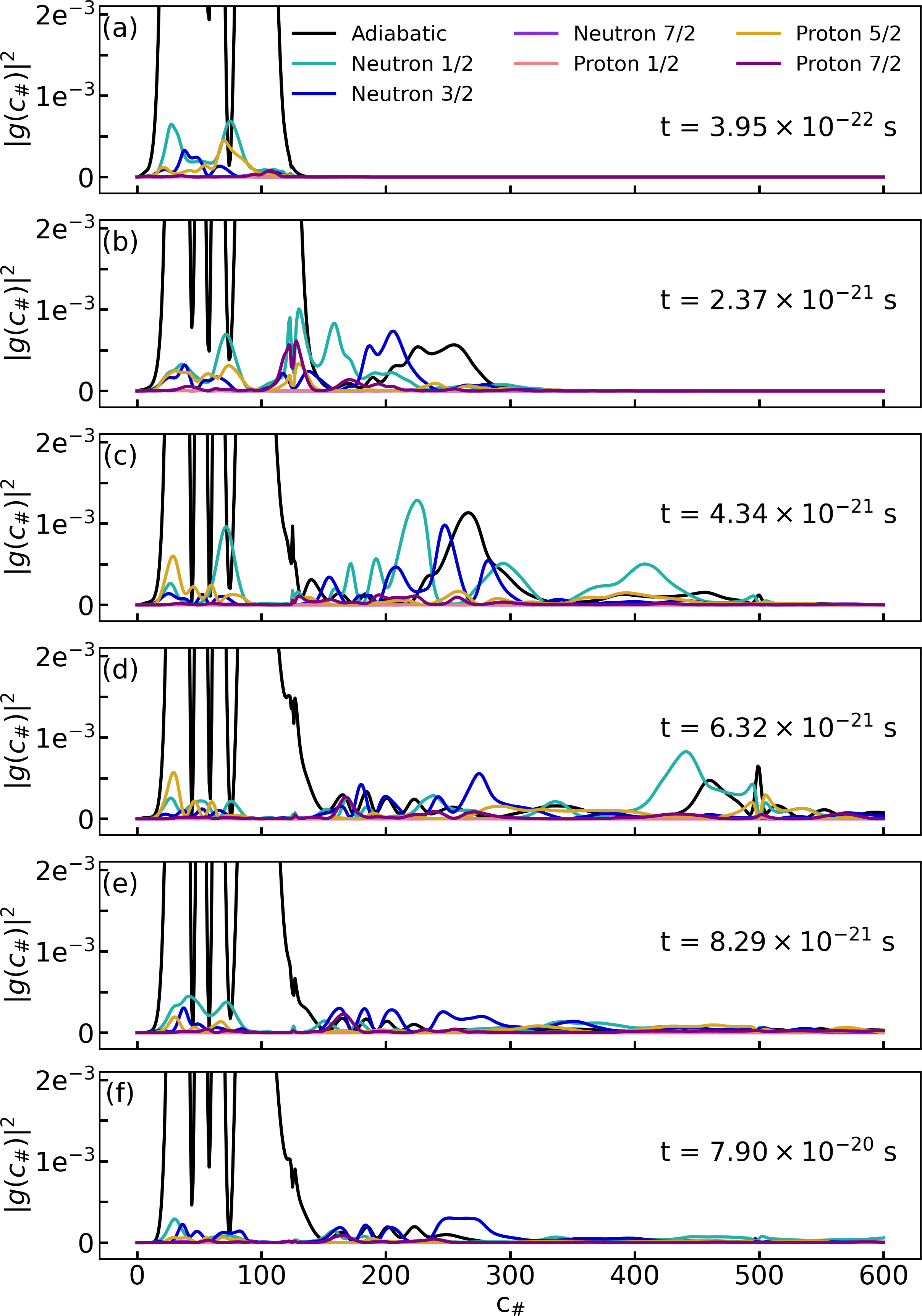}
\caption{Dynamical propagation of the wave packet with a relevant zoom. The different local squared norms of the wave function components $g_i$ are represented with respect to $c_\#$ at different times.}
\label{cfin_22}
\end{figure}

FIG. \ref{cfin_22} shows the evolution of the local probability densities $|g_i(c_\#,t)|^2$ at different times. Only the physical region ($c_\#\le600$) is displayed, the remaining interval corresponding to the absorbing layer. Panels (a)--(e) illustrate the propagation of the wave packet, whereas panel (f) shows the final state of the calculation.

At the end of the propagation, only $39.4\%$ of the initial norm remains inside the physical region, indicating that most of the wave packet has reached the absorbing layer. This behavior is consistent with the initial average energy being chosen at the top of the first fission barrier.

The propagation also reveals how the intrinsic composition of the wave packet evolves along the collective path. Beyond the saddle point ($c_\#\simeq267$), the adiabatic component progressively ceases to dominate and is overtaken near scission by the neutron $\Omega=1/2$ variational excitation. Conversely, inside the ground-state well, the different components remain localized and oscillate coherently. The residual probability leaking through the first barrier at late times is naturally interpreted as quantum tunneling.

\subsection{Probability fluxes and excited yields}

Probability fluxes are key quantities to characterize the dynamical evolution, as they provide access to the final probabilities associated with the different channels, thereby opening the way to the calculation of additional observables.

\subsubsection{Derivation of the probability fluxes}

The probability flux through a given position $c_s$ during a finite propagation time $t_f$ is defined as:
\begin{eqnarray}
\phi(c_s,t_f)
=
\int_0^{t_f} \mathrm{d}t \,
\frac{\mathrm{d}P(c_\#>c_s)}{\mathrm{d}t}(t),
\end{eqnarray}
where
\begin{eqnarray}
P(c_\#>c_s)(t)
=
\int_{c_s}^{800} \mathrm{d}c_\#
\sum_i |g_i(c_\#,t)|^2
\end{eqnarray}
is the probability contained beyond the position $c_s$ at time $t$.

The quantity $\phi(c_s,t_f)$ represents the net probability transfer through the point $c_s$ during the time interval $[0,t_f]$, including the possible sign changes associated with backward propagation. In general, this quantity does not converge for arbitrary values of $c_s$ and finite propagation times. However, due to the absorbing boundary conditions introduced in the SCIM Hamiltonian, any wave-packet component crossing the first barrier is eventually removed from the physical region. Therefore, for sufficiently large values of $c_s$, the total probability flux can be defined as:
\begin{eqnarray}\label{fluxequa}
\phi_{\mathrm{tot}}(c_s)
=
\int_0^{+\infty} \mathrm{d}t\,
\frac{\mathrm{d}P(c_\#>c_s)}{\mathrm{d}t}(t).
\end{eqnarray}

To obtain an explicit expression for $\phi_{\mathrm{tot}}$, we first develop the time derivative appearing in Eq.~(\ref{fluxequa}):
\begin{eqnarray}\label{cfin_43}
\displaystyle \frac{\text{d}\text{P}(c_\#>c_s) }{\text{d}t}(t) = \int_{c_s}^{800} \text{d}c_\# \; \sum_i \left( g_i(c_\#,t) \frac{\partial g^*_i(c_\#,t)}{\partial t} \right. \nonumber \\ \left. + g^*_i(c_\#,t)\frac{\partial g_i(c_\#,t)}{\partial t} \right). \qquad 
\end{eqnarray}

Using the collective-intrinsic Schrödinger equation, the integrand can be rewritten in the form:
\begin{eqnarray}\label{cfin_44}
\sum_i \left( g_i(c_\#,t) \frac{\partial g^*_i(c_\#,t)}{\partial t}  + g^*_i(c_\#,t)\frac{\partial g_i(c_\#,t)}{\partial t} \right) \nonumber \\ = -\frac{\partial J(c_\#,t)}{\partial c_\#}  \qquad \qquad,
\end{eqnarray}
which corresponds to the continuity equation, where $J(c_\#,t)$ is the associated probability current.

As demonstrated in Appendix~\ref{appendixa}, the probability current naturally separates into three contributions associated with the different terms of the SCIM Hamiltonian:
\begin{eqnarray}
\frac{\partial J(c_\#,t)}{\partial c_\#}
=
\frac{\partial J_V(c_\#,t)}{\partial c_\#}
 +\frac{\partial J_D(c_\#,t)}{\partial c_\#}
+\frac{\partial J_B(c_\#,t)}{\partial c_\#}. 
\end{eqnarray}
The contribution associated with the collective potential vanishes:
\begin{eqnarray}
J_V(c_\#,t)=0, \nonumber
\end{eqnarray}
whereas the dissipative and inertial contributions are given by:
\begin{eqnarray}
J_D(c_\#,t)
=
-\frac{2i}{\hbar}
\Im
\left(
\sum_{ij}
g_i(c_\#,t)
D_{ij}(c_\#)
g_j^*(c_\#,t)
\right), \nonumber 
\end{eqnarray}
and
\begin{eqnarray}
J_B(c_\#,t)
=
-\frac{8i}{\hbar}
\Im
\left(
\sum_{ij}
g_i(c_\#,t)
B_{ij}(c_\#)
\frac{\partial}{\partial c_\#}
g_j^*(c_\#,t)
\right). \nonumber 
\end{eqnarray}
Consequently, the total probability current reads:
\begin{eqnarray}
J(c_\#,t)
=
-\frac{2i}{\hbar}
\Im
\left(
\sum_{ij}
g_i(c_\#,t)
D_{ij}(c_\#)
g_j^*(c_\#,t)
\right)
\\\nonumber
-\frac{8i}{\hbar}
\Im
\left(
\sum_{ij}
g_i(c_\#,t)
B_{ij}(c_\#)
\frac{\partial}{\partial c_\#}
g_j^*(c_\#,t)
\right).
\end{eqnarray}
Inserting this expression into Eq.~(\ref{cfin_43}) gives:
\begin{eqnarray}
\frac{\mathrm{d}P(c_\#>c_s)}{\mathrm{d}t}(t)
=
-J(800,t)+J(c_s,t).
\end{eqnarray}
Because of the absorbing potential, the probability current vanishes at the end of the numerical domain. The total probability flux can therefore be directly evaluated at $c_s$:
\begin{widetext}
\begin{eqnarray}\label{express}
\phi_{tot}(c_s) = \int_{0}^{+\infty} \text{d}t \left[-\frac{2 i}{\hbar} \Im \left(\sum_{ij} g_i(c_s,t)D_{ij}(c_s)g^*_j(c_s,t)\right)  - \frac{8i}{\hbar}\Im \left(\sum_{ij}g_i(c_s,t)B_{ij}(c_\#)\frac{\partial}{\partial c_s}g_j^*(c_s,t)\right) \right] 
\end{eqnarray}
\end{widetext}

Finally, the total flux can be decomposed into contributions associated with the different channels:
\begin{eqnarray}\label{cfin_42}
\phi_{{tot}}(c_s)
=
\phi_0(c_s)
+
\sum_{i=1}^{N}
\phi_i(c_s),
\end{eqnarray}
where $N$ denotes the number of excited configurations included in the SCIM description.

\subsubsection{Flux analysis and excited yields}

\begin{figure}
\centering
\includegraphics[width=1.0\linewidth]{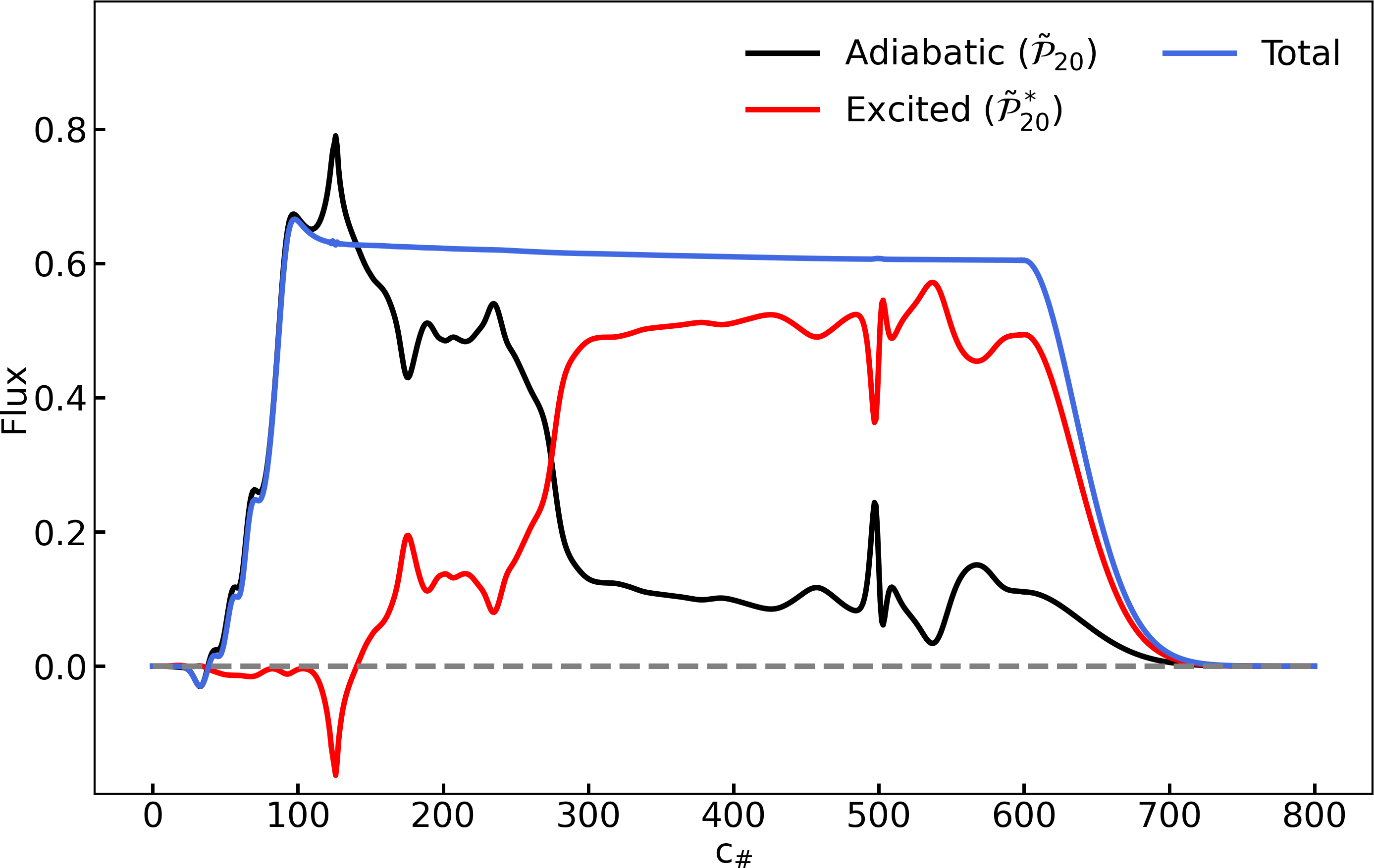}
\caption{Total, adiabatic and the excited probability fluxes after $t=7.90 \times 10^{-20}$s, and evaluated at different $c_\#$ values.}
\label{cfin_17}
\end{figure}

In FIG.~\ref{cfin_17}, we display the total probability flux (blue curve), together with its adiabatic (black curve) and excited (red curve) contributions. The excited contribution is defined as the sum of all $\phi_i$ with $i>0$. The fluxes are evaluated at the end of the propagation, $t=7.90\times10^{-20}$~s, for different values of the collective coordinate $c_\#$.

At first sight, the negative excited flux observed around $c_\#=120$ may appear surprising. However, this behavior can be understood by considering the different propagation paths contributing to the net flux. Some components initially belonging to the adiabatic channel propagate from the ground-state well toward the first barrier, thereby contributing positively to the adiabatic flux. They are then transferred to excited configurations and propagate back toward the ground-state well, thereby contributing negatively to the excited flux.

A remarkable feature is the rapid increase of the excited flux starting approximately at the saddle point ($c_\#=267$). Around $c_\#=280$, it already exceeds the adiabatic contribution. After this initial growth, the excited flux continues to increase gradually until reaching the scission region, located around $c_\#=495$, where both the adiabatic and excited contributions exhibit strong variations. We attribute these variations to the strong off-diagonal components of the SCIM Hamiltonian observed in this region.

The total flux, represented by the blue curve, provides additional information on the propagation and absorption processes. The plateau observed beyond the first barrier ($c_\#=120$) indicates that essentially all wave-function components crossing this region are eventually absorbed. At the end of the propagation, the total flux reaches a value of $0.606$ at $c_\#=600$, meaning that $60.6\%$ of the initial probability has been removed by the absorbing region. This result is consistent with the remaining norm of the propagated wave function, which was previously found to be $39.4\%$. The influence of the absorbing potential is visible beyond $c_\#=600$, where the flux progressively vanishes.

\begin{figure}
\centering
\includegraphics[width=1.0\linewidth]{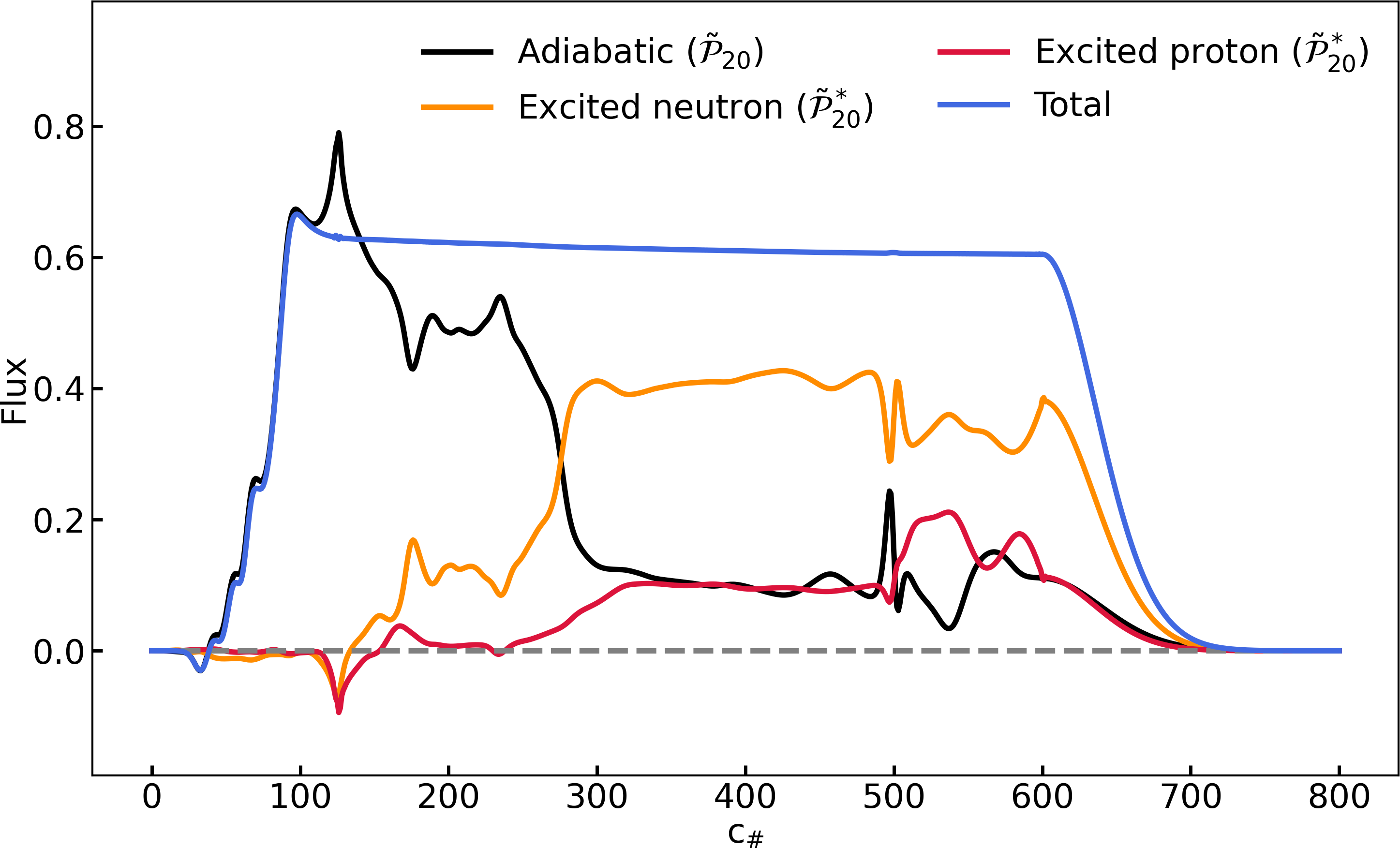}
\caption{Adiabatic and excited probability fluxes associated with both isospins, after $t=7.90 \times 10^{-20}$s, and evaluated at different $c_\#$ values.}
\label{cfin_15}
\end{figure}

In FIG. \ref{cfin_15}, we show the contributions of the different isospin channels to the excited flux, with neutron and proton excitations represented in orange and red, respectively. The adiabatic contribution is also indicated in black.
Within the present set of variational excitations, neutron excitations clearly dominate the excited flux. However, it remains difficult to determine whether this predominance reflects an intrinsic physical property of the fission dynamics or results from the specific selection of excited configurations considered in the present study.

An indication supporting a possible physical origin comes from the proton odd-even staggering observed at scission (around $c_\#=495$). This effect suggests that proton pairing correlations remain more robust against pair breaking than neutron ones in this region. Since the low-energy variational excitations considered here are mainly associated with pair-breaking mechanisms, the predominance of neutron excitations in the excited flux may therefore reflect genuine structural effects. Nevertheless, a more systematic study including a larger set of excited configurations would be required to fully assess this interpretation.

\begin{figure}
\centering
\includegraphics[width=1.0\linewidth]{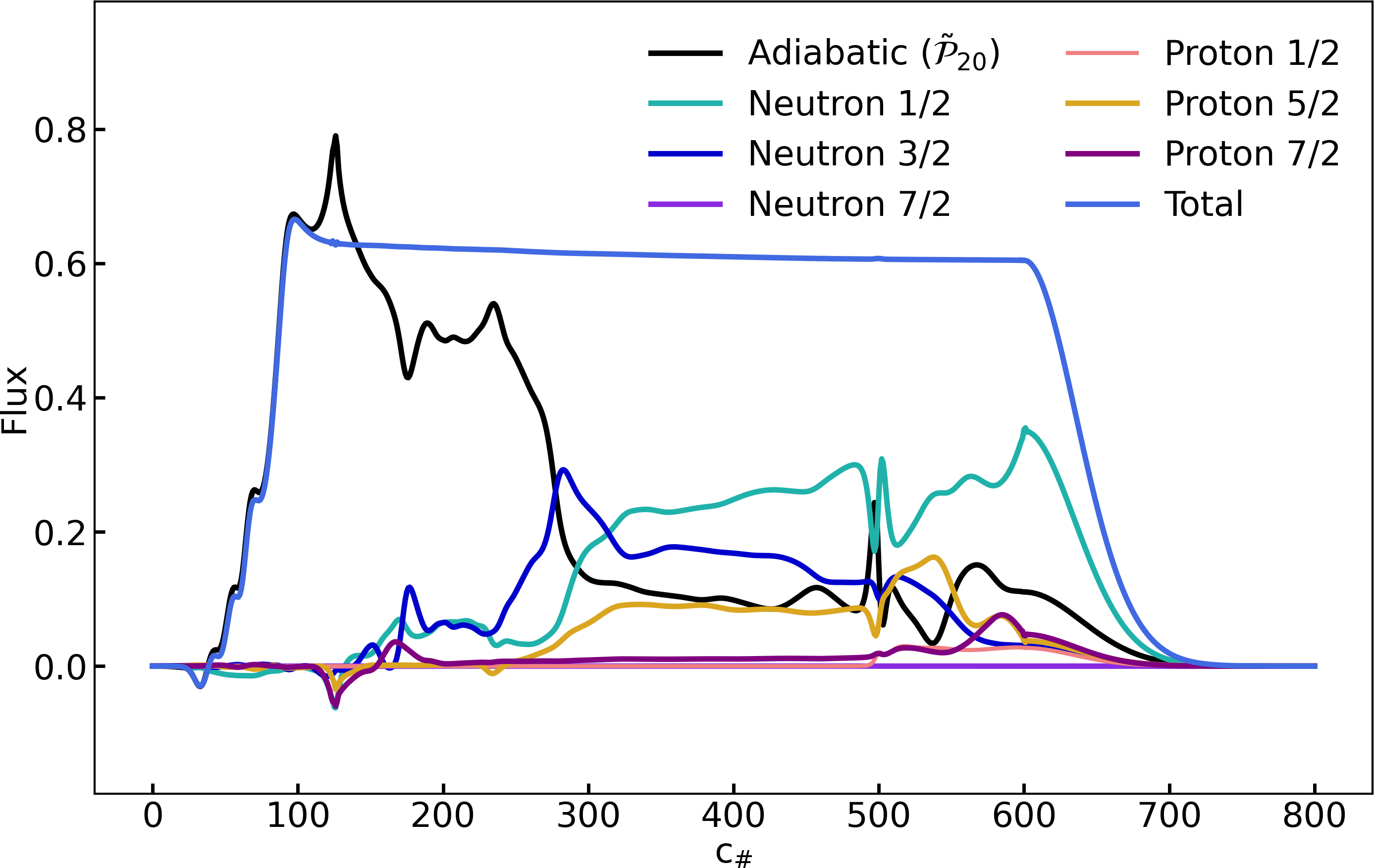}
\caption{Adiabatic and excited probability fluxes after $t=7.90 \times 10^{-20}$s, and evaluated at different $c_\#$ values.}
\label{cfin_14}
\end{figure}

To further analyze the excited flux, FIG. \ref{cfin_14} displays the adiabatic and individual excited probability fluxes at $t=7.90 \times 10^{-20}$~s, evaluated for different values of $c_\#$. This representation allows us to identify the respective contributions of the various excitations to the total excited flux.
As suggested by the previous analysis, the neutron $\Omega=1/2$ excitation provides the dominant contribution beyond $c_\#=315$. The neutron $\Omega=3/2$ and proton $\Omega=5/2$ excitations also contribute significantly to the excited flux. The contributions associated with the proton $\Omega=1/2$ and proton $\Omega=7/2$ excitations are smaller, but remain non-negligible in the vicinity of scission.

Conversely, no significant probability flux is observed for the neutron $\Omega=7/2$ excitation. Indeed, its contribution remains approximately three orders of magnitude smaller than those of the other excited states. This behavior is consistent with the weak $V_{SCIM}$, $B_{SCIM}$ and $D_{SCIM}$ couplings associated with this excitation, as discussed previously in Section \ref{dynaing}. \\

To conclude this analysis, we extract the excited yields associated with the different probability fluxes. The decomposition introduced in Eq.~(\ref{cfin_42}) provides direct access to the yields $\text{Y}_{i}(c_s)$ associated with the adiabatic and excited channels at a given scission coordinate $c_s$, defined as:
\begin{eqnarray}
\text{Y}_{i}(c_s) = \frac{\phi_{i}(c_s)}{\phi_{tot}(c_s)} .
\end{eqnarray}

Because of the strong flux fluctuations observed in the vicinity of scission, we evaluate the yields from the average flux values over the interval $445 \leq c_\# \leq 545$. The resulting yields are 15.8\% for the adiabatic component, 41.5\% for the neutron $\Omega=1/2$ excitation, 20.3\% for the neutron $\Omega=3/2$ excitation, 0.0\% for the neutron $\Omega=7/2$ excitation, 2.1\% for the proton $\Omega=1/2$ excitation, 17.5\% for the proton $\Omega=5/2$ excitation, and 2.8\% for the proton $\Omega=7/2$ excitation.

Overall, the excited channels account for 84.2\% of the total yield. This result clearly highlights the importance of including intrinsic excitations in the description of the fission dynamics.

\subsection{Fragment proton and neutron distributions}

The yields at the SCIM level are simply calculated as a weighted mixing of the $i$ various components of the SCIM wave-function, which represent both adiabatic and excited states. For the light ($l$) and heavy ($h$) fragments, they are expressed as:
\begin{eqnarray}\label{cfin_55}
 \text{Y}_{SCIM}(N_{l,h}) = \sum_{i=0}^{N} \text{Y}_i c^{2}_i(N_{l,h}),
\end{eqnarray}
and
\begin{eqnarray}
 \text{Y}_{SCIM}(Z_{l,h}) = \sum_{i=0}^{N} \text{Y}_i c^{2}_i(Z_{l,h}),
\end{eqnarray}
where $c^{2}_i(N_{l,h})$ and $c^{2}_i(Z_{l,h})$ denote the probabilities associated with the $i$-th component of the SCIM wave-function for producing a light or heavy fragment characterized by neutron number $N_{l,h}$ or proton number $Z_{l,h}$, respectively.

\begin{figure}
\centering
\includegraphics[width=1.0\linewidth]{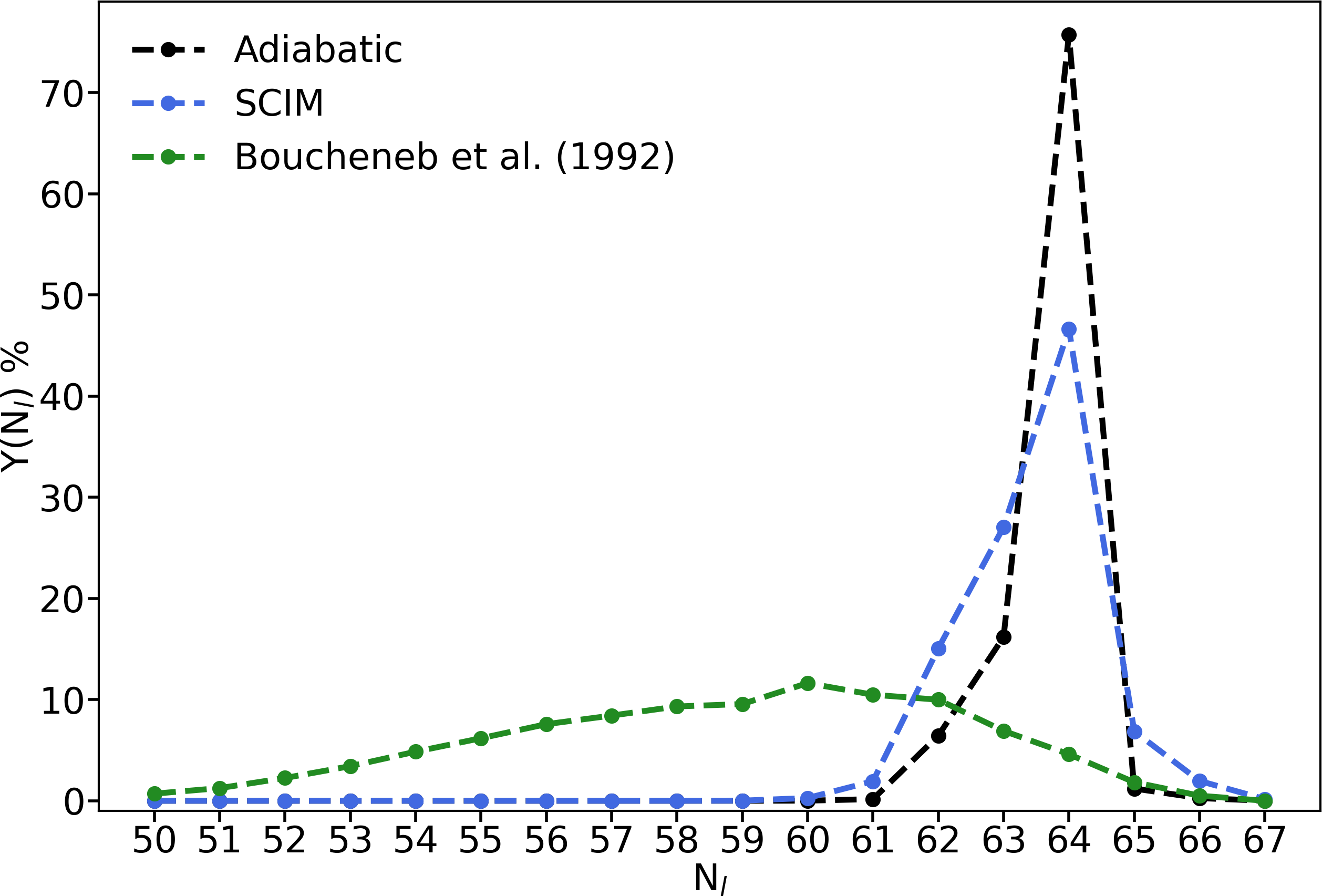}
\caption{Comparison between the adiabatic (black), the SCIM (blue), and the experimental (green) neutron yields extracted from Ref. \cite{Yield}.}
\label{cfin_19}
\end{figure}

In FIG. \ref{cfin_19}, we compare the adiabatic neutron yields (black curve) and the SCIM neutron yields (blue curve) evaluated at $c_\#=495$ with experimental data (green points) from Ref.~\cite{Yield}.
The inclusion of intrinsic excitations leads to a clear broadening of the SCIM neutron yields. This effect is particularly relevant since the adiabatic TDGCM calculations are known to generally underestimate the width of fragment distributions compared with experimental observations. Moreover, the odd-neutron contributions are significantly enhanced when excitations are included. This behavior is consistent with the pair-breaking nature of the variational excitations considered in the present study. The amplitude of the dominant fragmentation is also reduced, as expected from the redistribution of probability associated with the broadening of the distribution.

Nevertheless, both the adiabatic and SCIM neutron yields remain significantly different from the experimental distribution. This discrepancy is not unexpected, since the present calculation is restricted to a one-dimensional asymmetric collective path and does not include additional collective degrees of freedom. Extending the SCIM framework to multidimensional collective spaces will therefore be essential for more quantitative comparisons with experimental data.\\

\begin{figure}
\centering
\includegraphics[width=1.0\linewidth]{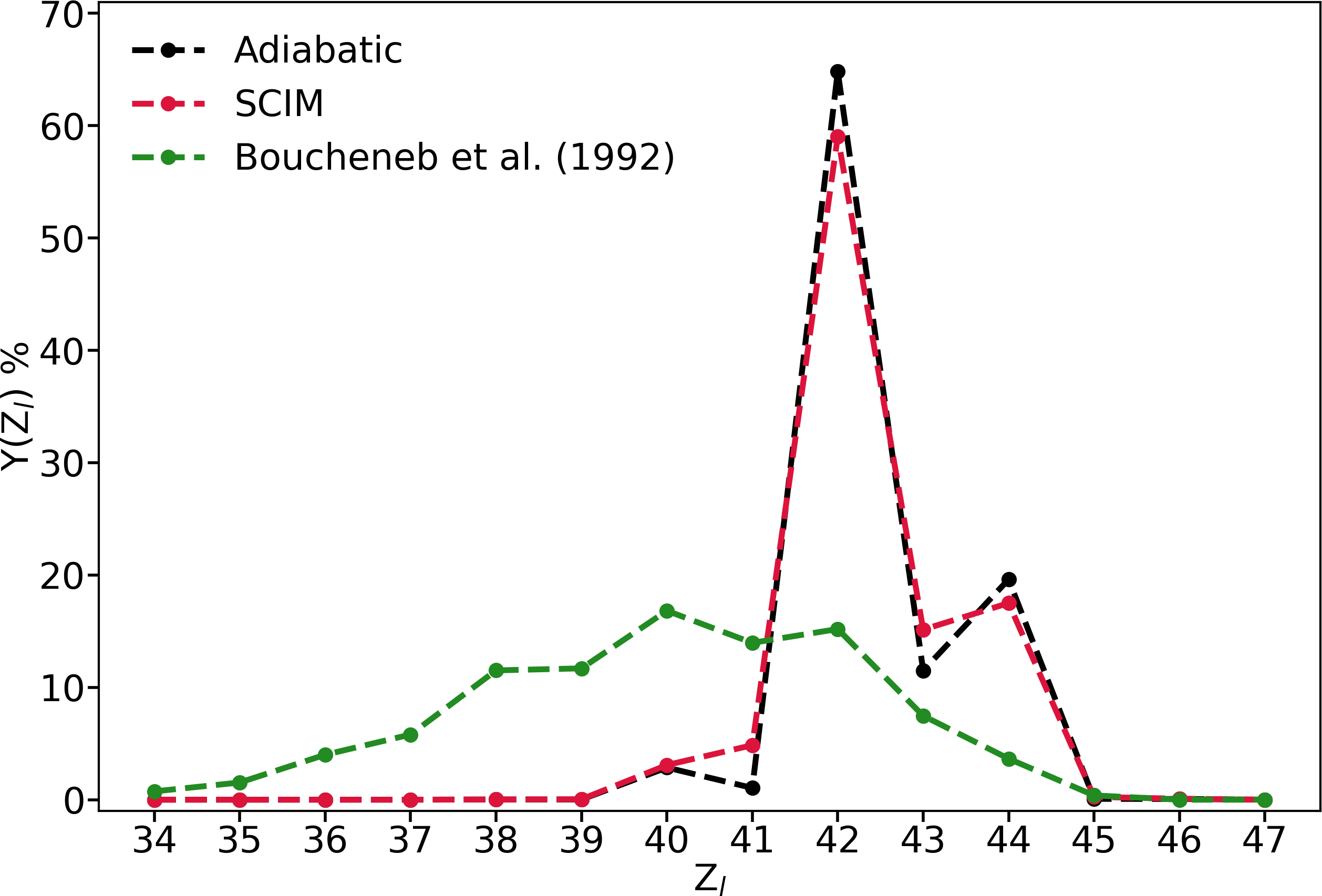}
\caption{Same as FIG. \ref{cfin_19} but for protons. Experimental data are extracted from Ref.\cite{Yield}.}
\label{cfin_20}
\end{figure}

In FIG. \ref{cfin_20}, we present the corresponding proton yields, comparing the adiabatic (black curve) and SCIM (red curve) results evaluated at $c_\#=495$ with experimental data (green points) from Ref.~\cite{Yield}.
The broadening induced by intrinsic excitations is also visible in the proton yields, although it is significantly weaker than for neutrons. This difference is consistent with the smaller total proton excited population obtained from the dynamical calculation. As discussed previously, whether this reduced proton contribution originates from genuine physical effects or from the restricted set of proton excitations considered here requires further investigation.

The modifications of the proton distributions also indicate the presence of pair-breaking effects, although their impact remains more moderate than in the neutron case. The SCIM proton yields provide an improvement compared with the purely adiabatic description, but significant deviations from the experimental distribution remain. As for the neutron yields, introducing additional collective coordinates in future multidimensional SCIM applications will be necessary to account for the full range of possible fragmentations.

\subsection{Dynamical energy balance at scission}

To conclude the analysis of the dynamical results, we investigate the energy balance at scission.
First, we evaluate the intrinsic excitation energy generated by the coupling between collective and intrinsic degrees of freedom.
Then, combining this quantity with the deformation and interaction energies at scission, we extract the pre-scission kinetic energy and the total kinetic energy. \\

To characterize the local intrinsic excitation energy along the collective path, we define:
\begin{eqnarray}\label{cfin_56}
 E^*(c_\#) = \sum_{i=1}^{6} 
 \frac{\phi_i(c_\#)}{\phi_{tot}(c_\#)}
 \left[
 E^{(i)}_{HFB}(c_\#)-E_{HFB}(c_\#)
 \right],
\end{eqnarray}
where $E^{(i)}_{HFB}(c_\#)$ corresponds to the HFB total binding energy associated with the $i$-th excitation evaluated at $c_\#$.

To obtain a representative value of the intrinsic excitation energy at scission, we consider the average of $E^*(c_\#)$ over the interval $445 \leq c_\# \leq 545$.
This procedure reduces the impact of the flux oscillations observed near scission and is consistent with the procedure used previously to extract the excited yields. The averaged intrinsic excitation energy, denoted $E_s^*$ is estimated as:
\begin{eqnarray}
E_s^*=7.55~\text{MeV}.
\end{eqnarray}
This value represents a significant amount of energy, corresponding approximately to the evaporation of one neutron after scission. \\

\begin{figure}
\centering
\includegraphics[width=1.0\linewidth]{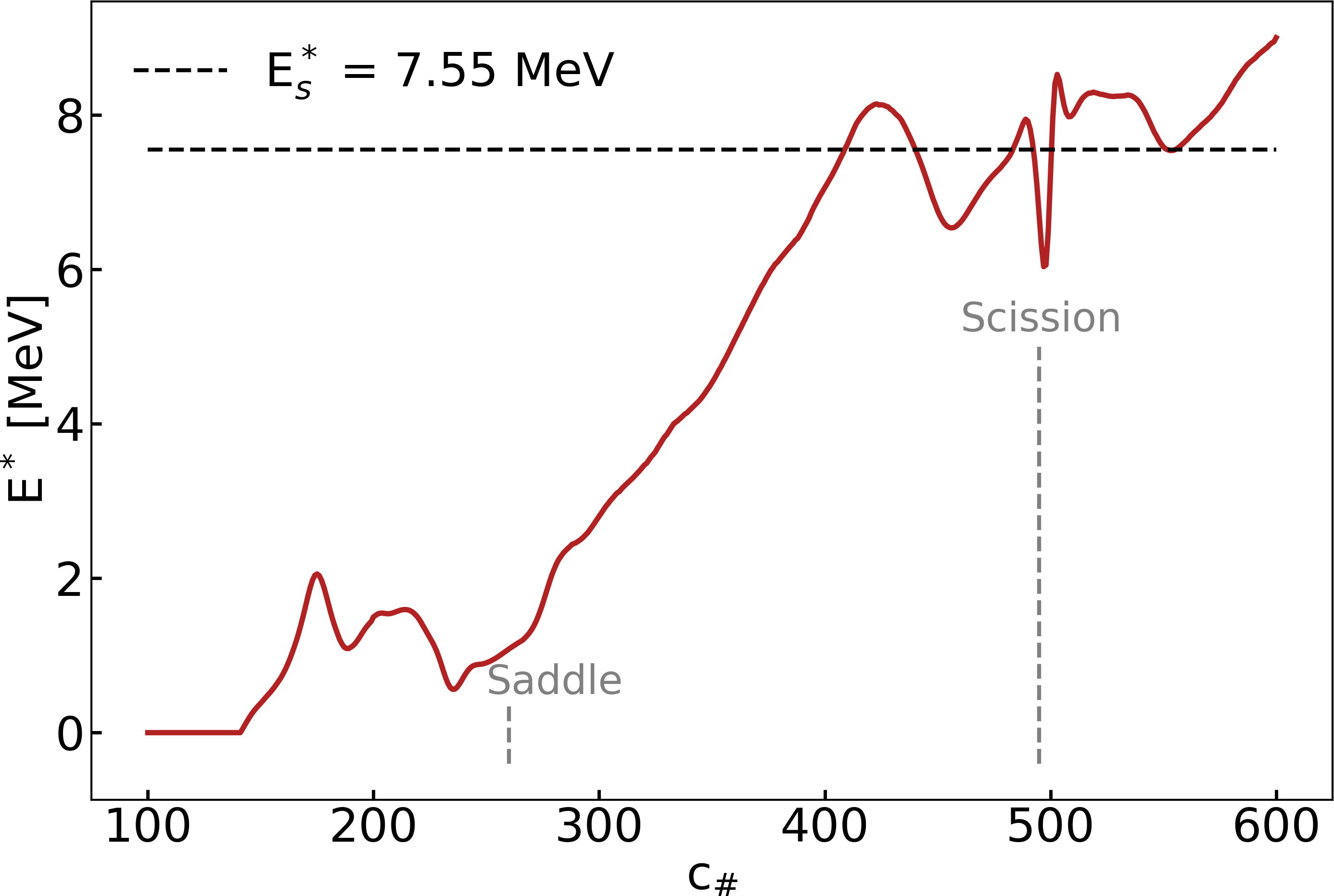}
\caption{Evolution of the local intrinsic excitation energy $E^*(c_\#)$ with respect to $c_\#$.}
\label{cfin_18}
\end{figure}

In FIG.~\ref{cfin_18}, we display the evolution of the local intrinsic excitation energy $E^*(c_\#)$. The averaged value at scission is also indicated.
The excitation process clearly starts around the saddle point ($c_\#=267$), where the different collective channels begin to couple.
Furthermore, the almost linear increase of $E^*(c_\#)$ between the saddle point and the scission region is particularly noticeable. Although the excited probability flux increases rapidly just after the saddle point and then more progressively, the corresponding intrinsic excitation energy continues to grow linearly because the energy carried by the excited components increases along the descent.

The approximately linear behavior of $E^*(c_\#)$ naturally motivates the definition of an effective dissipation coefficient:
\begin{eqnarray}
\gamma_{c_\#}^*
=
\frac{E^*(420)-E^*(267)}{420-267}
=
4.50\times10^{-2}~\text{MeV}.
\end{eqnarray}
This quantity provides an estimate of the amount of collective energy converted into intrinsic excitation per unit variation of the collective coordinate $c_\#$ beyond the saddle point.
Equivalently, it can be expressed with respect to the quadrupole moment $Q_{20}$:
\begin{eqnarray}
\gamma_{Q_{20}}^*
= 1.22\times10^{-3}~\text{MeV.fm}^{-2}.
\end{eqnarray}

Further investigations will be necessary to establish the physical relevance of these coefficients.
In particular, their values may depend on the nucleus, the initial conditions of the dynamics, and especially the average energy of the initial wave packet.
Moreover, the observed linear behavior may partly result from the specific set of intrinsic excitations considered in the present work.
Nevertheless, for a given nucleus and a fixed initial condition, such coefficients may provide useful qualitative estimates. \\

We now have all the necessary ingredients to establish the energy balance at scission.
At $c_\#=495$, the relevant quantities are:
the deformation energy of the fragments $E_D=26.85$~MeV,
the Coulomb interaction energy $E_C=178.743$~MeV,
the total interaction energy $E_{int}=152.50$~MeV,
and the energy difference between the top of the first barrier and scission,
$\Delta E_{1s}=33.31$~MeV. \\

The total excitation energy (TXE) is then obtained as:
\begin{eqnarray}
\text{TXE}=E_D+E_s^*=34.40~\text{MeV}.
\end{eqnarray}
This value is slightly larger than experimental estimates.
Indeed, the experimental neutron multiplicity for the
$^{239}$Pu($n_{th}$,f) reaction, combined with prompt gamma emission, corresponds to a TXE of approximately $30$~MeV \cite{nuBar}.
This discrepancy may originate from the deformation energy estimate, which is known to be particularly sensitive to the fragment separation procedure. \\

The pre-scission kinetic energy is defined as:
\begin{eqnarray}
E_{PS}=\Delta E_{1s}-E_s^*=25.76~\text{MeV}.
\end{eqnarray}

This quantity allows us to evaluate the total kinetic energy under the two hypotheses discussed in the first article of the trilogy \cite{trilogy1}.
In the first approach, only the Coulomb interaction energy is converted into fragment kinetic energy:
\begin{eqnarray}
\text{TKE}_C=E_{PS}+E_C=204.50~\text{MeV}.
\end{eqnarray}

In the second approach, the full interaction energy between fragments is considered:
\begin{eqnarray}
\text{TKE}_{int}=E_{PS}+E_{int}=178.26~\text{MeV}.
\end{eqnarray}

\begin{figure}
\centering
\includegraphics[width=1.0\linewidth]{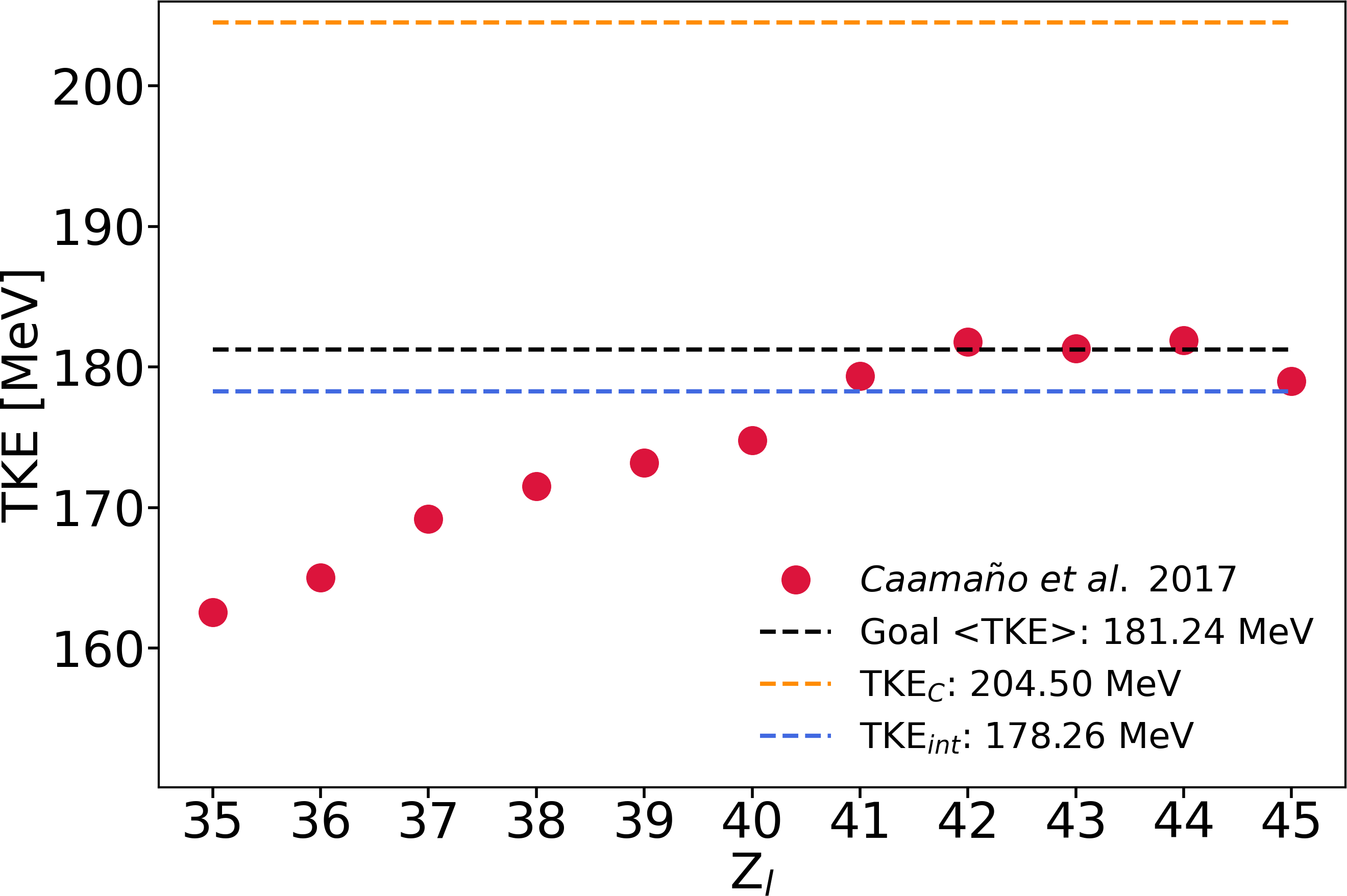}
\caption{Average experimental TKE associated with the $^{239}$Pu($n_{th}$,f) reaction, displayed with respect to the light fragment proton number $Z_l$ (filled circles) \cite{TKEPu}. Two different lines correspond to the different quantities, TKE$_C$ and TKE$_{int}$, discussed in the text.}
\label{cfin_23}
\end{figure}

To compare these values with experimental data, FIG.~\ref{cfin_23} displays the average experimental TKE for the
$^{239}$Pu($n_{th}$,f) reaction as a function of the light fragment proton number $Z_l$ \cite{TKEPu}.
The expected TKE extracted from the SCIM fragment distribution at scission, referred to as the \enquote{goal TKE}, is also shown.
It corresponds to $181.24$~MeV.
The values of $\text{TKE}_C$ and $\text{TKE}_{int}$ are indicated for comparison.

The TKE obtained using the total interaction energy is significantly closer to the \enquote{goal TKE} than the value obtained from the Coulomb interaction alone.
This result indicates that the nuclear interaction energy cannot be neglected in the evaluation of the scission energy balance.
However, this conclusion should be considered with caution, as the different contributions entering the energy balance may depend on the fragment separation procedure.
Further investigations will therefore be required.

Finally, the pre-scission kinetic energy represents $14.4\%$ of $\text{TKE}_{int}$.
This order of magnitude is consistent with the values reported in Refs.~\cite{LacMet}, obtained using a stochastic mean-field approach to explore the nuclear phase space.

\section{Conclusion and perspectives}\label{conclu}

In this third article of the trilogy, we have presented the first dynamical application of the SCIM approach.
The main objective was to extend the static description developed in the previous works towards a fully time-dependent treatment including intrinsic excitations.

First, we have addressed the regularization of the dynamical ingredients entering the SCIM Hamiltonian, both for adiabatic and excited configurations.
This procedure allows us to focus on the relevant second-order collective dynamics described by the collective potential $V_{SCIM}$, the collective inertia tensor $B_{SCIM}$, and the collective dissipative tensor $D_{SCIM}$.
For the adiabatic path, a formal and numerical comparison between the GOA and the SCIM approaches has been performed, showing the global consistency between both descriptions in the appropriate limit.
For excited configurations, we have analyzed the diagonal and off-diagonal properties of the collective ingredients, highlighting the importance of the coupling terms and, in particular, the dominant role played by the pure neutron and proton channels.

Second, we have detailed the numerical solution of the collective-intrinsic Schr\"odinger equation.
This allowed us to extract the probability fluxes associated with both the adiabatic and excited components of the evolving wave packet.
From these quantities, we have obtained the excited yields at scission and demonstrated the significant contribution of intrinsic excitations to the fission dynamics.

Finally, we have calculated the proton and neutron fragment distributions at the SCIM level and shown that the inclusion of intrinsic excitations broadens the predicted yields and increases the odd components associated with pair-breaking effects.
Moreover, we have established a dynamical energy balance at scission by introducing, for the first time in this type of TDGCM approach, an intrinsic excitation energy.
The resulting analysis provides new insight into the interplay between collective motion, intrinsic excitation, and fragment observables. \\

The work presented throughout this trilogy should be considered as a first step toward fully multidimensional SCIM applications.
Beyond the challenges associated with the extension to two collective coordinates, several open questions remain regarding the extraction of observables at scission.
In particular, developing a unified framework to determine fragment observables consistently for both adiabatic and excited configurations would be highly desirable.
Such a framework should provide a simultaneous description of fragment proton and neutron distributions, angular momentum distributions, and the energy balance at scission.

Furthermore, a deeper understanding of the fragment separation mechanism at scission remains essential.
Clarifying this aspect will be crucial to reduce the uncertainties associated with the extraction of observables and to fully exploit the predictive capabilities of the SCIM approach.
Work in this direction is currently in progress.\\ \\ 

\noindent \textbf{Acknowledgment}: N.P and P.C. would like to thank J.F. Berger
for his kindness throughout this work. N.P. dedicates this first application 
of the SCIM approach to the memory of D. Gogny. The work of W.Y. was supported by the U.S. Department of Energy, Office of Science, Office of Nuclear Physics under the contract No.
DE-AC02- 05CH11231 (LBNL). The work of L.M.R. is supported by Spanish Agencia Estatal de Investigacion 
(AEI) of the Ministry of Science and Innovation under Grant No. 
PID2024-159559NB-C21.

\appendix

\section{Probability current}\label{appendixa}

To find the expression of the probability current $J$, we start by rephrasing the derivatives in Eq.(\ref{cfin_43}) thanks to the collective-intrinsic Schr\"odinger equation. We find
\begin{eqnarray}\label{cfin_45}
\frac{\partial g_i(c_\#,t)}{\partial t}  = -\frac{i}{\hbar} \sum_j \left( \mathcal{H}_{SCIM} \right)_{ij}(c_\#) g_j(c_\#,t),
\end{eqnarray}
and
\begin{eqnarray}\label{cfin_46}
\frac{\partial g^*_i(c_\#,t)}{\partial t}  = \frac{i}{\hbar} \sum_j \left( \mathcal{H}_{SCIM}\right)_{ij}(c_\#) g^*_j(c_\#,t).
\end{eqnarray}
\noindent Inserting Eq.(\ref{cfin_45}) and Eq.(\ref{cfin_46}) into Eq.(\ref{cfin_44}), one obtains:
\begin{widetext}
\begin{eqnarray}\label{cfin_47}
 -\frac{\partial J(c_\#,t)}{\partial c_\#}  = \frac{i}{\hbar}\sum_{ij}\left[g_i(c_\#,t) \left(V_{ij}(c_\#)  + \left[D_{ij}(c_\#)\frac{\partial}{\partial c_\#}\right]^{(1)} + \left[B_{ij}(c_\#)\frac{\partial}{\partial c_\#}\right]^{(2)}\right) g^*_j(c_\#,t)
 \right. \nonumber \\ \left. - g^*_i(c_\#,t) \left(V_{ij}(c_\#)  + \left[D_{ij}(c_\#)\frac{\partial}{\partial c_\#}\right]^{(1)} + \left[B_{ij}(c_\#)\frac{\partial}{\partial c_\#}\right]^{(2)}\right)g_j(c_\#,t) \right]. 
\end{eqnarray}
\end{widetext}
Then, we separate Eq.(\ref{cfin_47}) into three parts:
\begin{eqnarray}
 \frac{\partial J(c_\#,t)}{\partial c_\#} = \frac{\partial J_V(c_\#,t)}{\partial c_\#}  +\frac{\partial J_D(c_\#,t)}{\partial c_\#} \nonumber \\  + \frac{\partial J_B(c_\#,t)}{\partial c_\#}, 
\end{eqnarray}
with:
\begin{eqnarray}
-\frac{\partial J_V(c_\#,t)}{\partial c_\#}  = \frac{i}{\hbar}\sum_{ij}\left[g_i(c_\#,t) V_{ij}(c_\#)  g^*_j(c_\#,t) \right. \nonumber \\
\left.  - g^*_i(c_\#,t)V_{ij}(c_\#)g_j(c_\#,t)\right], \nonumber
\end{eqnarray}
\begin{eqnarray}
-\frac{\partial J_D(c_\#,t)}{\partial c_\#}  = \frac{i}{\hbar}\sum_{ij}\left[ g_i(c_\#,t) \left[D_{ij}(c_\#)\frac{\partial}{\partial c_\#}\right]^{(1)}  g^*_j(c_\#,t) \right. \nonumber
\\ \left. - g^*_i(c_\#,t)\left[D_{ij}(c_\#)\frac{\partial}{\partial c_\#}\right]^{(1)}(c_\#)g_j(c_\#,t)\right], \qquad  \qquad \nonumber
\end{eqnarray}
and
\begin{eqnarray}
-\frac{\partial J_B(c_\#,t)}{\partial c_\#}  = \frac{i}{\hbar}\sum_{ij} \left[g_i(c_\#,t) \left[B_{ij}(c_\#)\frac{\partial}{\partial c_\#}\right]^{(2)}  g^*_j(c_\#,t) \right.  \nonumber 
\\ \left. \left. - g^*_i(c_\#,t)[B_{ij}(c_\#)\frac{\partial}{\partial c_\#}\right]^{(2)}(c_\#)g_j(c_\#,t)\right]. \qquad \qquad \nonumber
\end{eqnarray}

For the calculation of $J_V$, we first remark that the symmetry of the SCIM potential $V$ directly leads to:
\begin{eqnarray}\label{cfin_51}
J_V(c_\#,t) = 0.
\end{eqnarray}
In Eq.(\ref{cfin_51}), we have set the constant part to zero, as it is irrelevant in the subsequent analysis. \\

\noindent To determine $J_D$, one sees that the quantity $-\frac{\partial J_D(c_\#,t)}{\partial c_\#} $ explicitly reads as:
\begin{eqnarray}\label{cfin_50}
-\frac{\partial  J_D(c_\#,t)}{\partial c_\#} = \frac{i}{\hbar}\sum_i \qquad \qquad \qquad \qquad \\ \times \left[g_i(c_\#,t) \sum_j \left(\frac{\partial}{\partial c_\#}D_{ij}(c_\#) + D_{ij}(c_\#)  \frac{\partial}{\partial c_\#}\right)g^*_j(c_\#,t)
\right. \nonumber \\ \left. \nonumber - g^*_i(c_\#,t) \sum_j \left(\frac{\partial}{\partial c_\#}D_{ij}(c_\#) + D_{ij}(c_\#)  \frac{\partial}{\partial c_\#}\right) g_j(c_\#,t)\right].
\end{eqnarray}
Then, we note that:
\begin{eqnarray}\label{cfin_48}
\sum_{ij} g_i(c_\#,t)\frac{\partial}{\partial c_\#} \left(D_{ij}(c_\#)g^*_j(c_\#,t)\right) = \qquad \qquad \nonumber \\ \nonumber
\frac{\partial}{\partial c_\#} \left(\sum_{ij} g_i(c_\#,t)D_{ij}(c_\#)g^*_j(c_\#,t) \right) \\  - \sum_{ij} \frac{\partial}{\partial c_\#} \left(g_i(c_\#,t)\right)D_{ij}(c_\#)g^*_j(c_\#,t),
\end{eqnarray}
and similarly:
\begin{eqnarray}\label{cfin_49}
- \sum_{ij} g^*_i(c_\#,t)\frac{\partial}{\partial c_\#} \left(D_{ij}(c_\#)g_j(c_\#,t)\right) = \qquad \qquad \nonumber \\ \nonumber
- \frac{\partial}{\partial c_\#} \left(\sum_{ij} g^*_i(c_\#,t)D_{ij}(c_\#)g_j(c_\#,t) \right) \\  + \sum_{ij} \frac{\partial}{\partial c_\#} \left( g^*_i(c_\#,t) \right)D_{ij}(c_\#)g_j(c_\#,t).
\end{eqnarray}
Inserting Eqs.(\ref{cfin_48}) and (\ref{cfin_49}) into Eq.(\ref{cfin_50}) and using the skew-symmetry property of the dissipation tensor $D$ leads to:
\begin{eqnarray}
-\frac{\partial J_D(c_\#,t)}{\partial c_\#}  = \frac{i}{\hbar}\frac{\partial}{\partial c_\#} (\sum_{ij} g_i(c_\#,t)D_{ij}(c_\#)g^*_j(c_\#,t)). \nonumber \\ \left.
- g^*_i(c_\#,t)D_{ij}(c_\#)g_j(c_\#,t) \right) \nonumber
\end{eqnarray}
Consequently, $J_D(c_\#,t)$ has for expression:
\begin{eqnarray}
J_D(c_\#,t) = -\frac{2 i}{\hbar} \Im \left(\sum_{ij} g_i(c_\#,t)D_{ij}(c_\#)g^*_j(c_\#,t)\right). \nonumber
\end{eqnarray}

\noindent To evaluate $J_B$, one starts with the quantity $-\frac{\partial J_B(c_\#,t)}{\partial c_\#} $ that can be expressed as:
\begin{widetext}
\begin{eqnarray}
-\frac{\partial J_B(c_\#,t)}{\partial c_\#}  = \frac{i}{\hbar}\sum_i \left[g_i(c_\#,t) \sum_j \left( \frac{\partial^2}{\partial c^2_\#}B_{ij}(c_\#) + 2 \frac{\partial}{\partial c_\#}B_{ij}(c_\#)\frac{\partial}{\partial c_\#} +  B_{ij}(c_\#)  \frac{\partial^2}{\partial c^2_\#} \right) g^*_j(c_\#,t)
\right. \nonumber \\  \left. - g^*_i(c_\#,t) \sum_j \left(\frac{\partial^2}{\partial c^2_\#}B_{ij}(c_\#) + 2 \frac{\partial}{\partial c_\#}B_{ij}(c_\#)\frac{\partial}{\partial c_\#} +  B_{ij}(c_\#)  \frac{\partial^2}{\partial c^2_\#}\right)g_j(c_\#,t) \right].
\end{eqnarray}
\end{widetext}
We remark that:
\begin{eqnarray}\label{cfin_52}
\sum_{ij} g_i(c_\#,t) \frac{\partial^2}{\partial c^2_\#}B_{ij}(c_\#) g^*_j(c_\#,t)
= \qquad \qquad \nonumber \\ \nonumber \frac{\partial^2}{\partial c^2_\#} \left( \sum_{ij} g_i(c_\#,t)B_{ij}(c_\#) g^*_j(c_\#,t)\right) 
\\ \nonumber + \sum_{ij} \frac{\partial^2}{\partial c^2_\#}\left(g_i(c_\#,t)\right)B_{ij}(c_\#) g^*_j(c_\#,t) \\
 - 2 \frac{\partial}{\partial c_\#}\left(\sum_{ij} \frac{\partial}{\partial c_\#}(g_i(c_\#,t)\right) B_{ij}(c_\#) g^*_j(c_\#,t)),
\end{eqnarray}
and similarly:
\begin{eqnarray}
- \sum_{ij} g^*_i(c_\#,t) \frac{\partial^2}{\partial c^2_\#}B_{ij}(c_\#) g_j(c_\#,t)
= \qquad \qquad \nonumber \\ \nonumber -\frac{\partial^2}{\partial c^2_\#}\left(\sum_{ij} g^*_i(c_\#,t)B_{ij}(c_\#) g_j(c_\#,t) \right)
\\ \nonumber - \sum_{ij} \frac{\partial^2}{\partial c^2_\#}\left(g^*_i(c_\#,t)\right)B_{ij}(c_\#) g_j(c_\#,t) \nonumber\\
 + 2 \frac{\partial}{\partial c_\#}\left(\sum_{ij} \frac{\partial}{\partial c_\#}(g^*_i(c_\#,t)\right)B_{ij}(c_\#) g_j(c_\#,t)). 
\end{eqnarray}
In addition, the following property holds:
\begin{eqnarray}
2 \sum_{ij} g_i(c_\#,t) \frac{\partial}{\partial c_\#}(B_{ij}(c_\#)\frac{\partial}{\partial c_\#} g^*_j(c_\#,t)) = \qquad \qquad \nonumber \\ 2 \frac{\partial}{\partial c_\#} \left(\sum_{ij} g_i(c_\#,t) B_{ij}(c_\#)\frac{\partial}{\partial c_\#} g^*_j(c_\#,t)) \nonumber \right. \qquad
\\ \left.  - 2 \sum_{ij}\frac{\partial}{\partial c_\#}(g_i(c_\#,t)\right)B_{ij}(c_\#)\frac{\partial}{\partial c_\#} g^*_j(c_\#,t), \qquad
\end{eqnarray}
and similarly:
\begin{eqnarray}\label{cfin_53}
-2 \sum_{ij} g^*_i(c_\#,t) \frac{\partial}{\partial c_\#}(B_{ij}(c_\#)\frac{\partial}{\partial c_\#} g^*_j(c_\#,t)) = \qquad \qquad \nonumber \\ -2 \frac{\partial}{\partial c_\#}\left(\sum_{ij} g^*_i(c_\#,t) B_{ij}(c_\#)\frac{\partial}{\partial c_\#} g_j(c_\#,t)\right) \qquad \nonumber
\\  + 2 \sum_{ij}\frac{\partial}{\partial c_\#}\left(g^*_i(c_\#,t)\right)B_{ij}(c_\#)\frac{\partial}{\partial c_\#} g_j(c_\#,t). \qquad
\end{eqnarray}
Using Eqs.(\ref{cfin_52}) to (\ref{cfin_53}) along with the symmetry of the inertia tensor $B$ provides:
\begin{eqnarray}
-\frac{\partial J_B(c_\#,t)}{\partial c_\#}  = \frac{4i}{\hbar}\frac{\partial}{\partial c_\#}\left(\sum_{ij}g_i(c_\#,t)B_{ij}(c_\#)\frac{\partial}{\partial c_\#}g_j^* \right. \nonumber \\ \left. - g^*_i(c_\#,t)B_{ij}(c_\#)\frac{\partial}{\partial c_\#}g_j(c_\#,t) \right). \qquad
\end{eqnarray}
By identification, $J_B(c_\#,t)$ has for expression:
\begin{eqnarray}
J_B(c_\#,t)= - \frac{8i}{\hbar}\Im \left(\sum_{ij}g_i(c_\#,t)B_{ij}(c_\#)\frac{\partial}{\partial c_\#}g_j^*(c_\#,t)\right). \nonumber
\end{eqnarray}
Accordingly, the probability current $J$ reads as:
\begin{eqnarray}
J(c_\#,t) = -\frac{2 i}{\hbar} \Im \left(\sum_{ij} g_i(c_\#,t)D_{ij}(c_\#)g^*_j(c_\#,t)\right) \nonumber
 \\ - \frac{8i}{\hbar}\Im \left(\sum_{ij}g_i(c_\#,t)B_{ij}(c_\#)\frac{\partial}{\partial c_\#}g_j^*(c_\#,t)\right).
\end{eqnarray}

\bibstyle{apsrev4-2}

\bibliography{dynamics_SCIM}

\end{document}